%% file: main.tex
\documentclass[twocolumn]{aastex631}

\newcommand{\dmunits}{${\rm pc}\;{\rm cm}^{-3}$}
\newcommand{\rmunits}{${\rm rad}\;{\rm m}^{-2}$}
\newcommand{\swiftj}{Swift J1818.0--1607 }

\graphicspath{{./}{figures/}}

\usepackage{natbib}
\usepackage{url}

\submitjournal{ApJ}

\shorttitle{Swift J1818.0--1607 Radio Observations}
\shortauthors{Lewis et al.}

\begin{document}

\title{Multifrequency Radio Observations of the Magnetar Swift J1818.0--1607}

\correspondingauthor{Evan F.~Lewis}
\email{efl0003@mix.wvu.edu}

\author[0000-0002-2972-522X]{Evan F.~Lewis}
\affiliation{West Virginia University, Department of Physics and Astronomy, P. O. Box 6315, Morgantown, WV 26506, USA}
\affiliation{Center for Gravitational Waves and Cosmology, West Virginia University, Chestnut Ridge Research Building, Morgantown, WV 26506, USA}

\author[0000-0003-4046-884X]{Harsha Blumer}
\affiliation{West Virginia University, Department of Physics and Astronomy, P. O. Box 6315, Morgantown, WV 26506, USA}
\affiliation{Center for Gravitational Waves and Cosmology, West Virginia University, Chestnut Ridge Research Building, Morgantown, WV 26506, USA}

\author[0000-0001-5229-7430]{Ryan S.~Lynch}
\affiliation{Green Bank Observatory, P.O. Box 2, Green Bank, WV 24944, USA}

\author[0000-0001-7697-7422]{Maura A.~McLaughlin}
\affiliation{West Virginia University, Department of Physics and Astronomy, P. O. Box 6315, Morgantown, WV 26506, USA}
\affiliation{Center for Gravitational Waves and Cosmology, West Virginia University, Chestnut Ridge Research Building, Morgantown, WV 26506, USA}

\begin{abstract}

We report on Green Bank Telescope observations of the radio magnetar Swift J1818.0--1607 between 820 MHz and 35 GHz, taken from six to nine months after its 2020 March outburst. We obtained multi-hour observations at six frequencies, recording polarimetric, spectral, and single-pulse information. The spectrum peaks at a frequency of $5.4 \pm 0.6$ GHz, making Swift J1818.0--1607 one of many radio magnetars which exhibit a gigahertz-peaked spectrum (GPS).  The radio flux decays steeply above the peak frequency, with in-band spectral indices $\alpha < -2.3 $ above 9 GHz. The emission is highly ($> 50\%$) linearly polarized, with a lower degree ($< 30\%$) of circular polarization which can change handedness between single pulses. Across the frequency range of our observations, the time-integrated radio profiles share a common shape: a narrow ``pulsar-like'' central component flanked by ``magnetar-like'' components comprised of bright, spiky subpulses. The outer profile components exhibit larger degrees of flux modulation when compared to the central pulse component. 

\end{abstract}

\section{Introduction} \label{sec:intro}

Magnetars are a subclass of neutron stars characterized by their large inferred surface magnetic field strengths ($> 10^{14}$ G) and long rotation periods ($> 1$ s) \citep{kb17}. Most magnetars are discovered through their X-ray and gamma-ray outbursts, with burst luminosities exceeding the spin-down luminosity available from the conversion of rotational angular momentum, which is the energy source for the majority of pulsars. Magnetar emission is instead likely powered by the decay of these extreme internal magnetic fields \citep{dt92}.

Pulsed radio emission has only been detected from six magnetars\footnote{\url{https://www.physics.mcgill.ca/~pulsar/magnetar/main.html}} including the most recent discovery, Swift J1818.0--1607. In the radio band, magnetar emission tends to be highly linearly polarized, with a smaller, but nonzero, degree of circular polarization (\citealt{ksj+07,crj+08,ljl+24}). The polarization position angle (PPA) of the emission often follows the S-shaped curve predicted by the rotating vector model (RVM) \citep{rvm_rc}, though many deviations from this model have been observed as well as changes in the sign of the slope of the PPA (\citealt{dlb+19,ljs+21,ljl+24}). The emission is composed largely of short, spiky `sub-pulses' falling within a larger pulse envelope \citep{lbb+12}. Magnetar emission can possess flat or even inverted spectral indices over large intervals of the radio spectrum, as in the case of XTE J1810--197 (\citealt{eta+21,tbb+22}) and the Galactic center magnetar SGR J1745--2900 (\citealt{tek+15, tde+17}). This is at odds with the majority of rotation-powered pulsars, whose flux follows power-law spectra with an average spectral index of $-$1.6 \citep{jvk+18}.

A unifying feature in magnetar emission across all bands is its temporal variability; particularly in the initial days to weeks after the onset of an outburst, there can be drastic changes in the shape and polarization properties of the integrated radio pulse profile, the overall shape of the spectrum, and single-pulse properties of the radio emission (e.g. \citealt{crh+16,scs+17, pmp+18, lys+23}). Large degrees of variability in the spin-down rate of the star are common, and magnetars also exhibit discrete glitching events and pulse profile changes suggestive of a rapidly changing magnetosphere (\citealt{ljs+21,rsl+22}). Magnetars have also been suggested as a central engine powering some extragalactic fast radio bursts (FRBs), particularly following the detection of FRB-like bursts from the Galactic magnetar SGR 1935+2154 (\citealt{chime_frb_mag,brb+20}). The underlying mechanisms of radio magnetar emission can provide valuable information to understand the FRB population as well as the properties of the magnetospheres surrounding neutron stars.

\swiftj (hereafter J1818) was discovered on 12 March 2020 when the \textit{Swift}-Burst Alert Telescope detected a 0.01-s duration gamma-ray burst (\citealt{2020GCN.27373....1E,2020GCN.27384....1S}). A 1.36-s periodicity was subsequently detected with NICER \citep{2020ATel13551....1E}. Within two days of the X-ray outburst, observations with the Lovell and Effelsberg radio telescopes yielded coherent radio pulsations, making J1818 the sixth known radio-loud magnetar as well as the fastest-spinning member of the magnetar population (\citealt{2020ATel13553....1K,2020ATel13554....1R}). 

Like other radio magnetars, J1818 exhibits changes in the shape and number of components in its integrated profile, high degrees of linear polarization, a variable spin-down rate, and flux variability across a range of timescales (\citealt{lsj+20,hys+21,rsl+22,fbr+24}). Its X-ray properties post-outburst were also reminiscent of other magnetar outbursts (\citealt{bs20,hbg+20}). More recently, the radio flux of J1818 increased from 2022 December to 2023 March, but subsequently decayed, and no detections have been published since then, suggesting J1818 may have entered a quiescent period after its outburst \citep{dld+24}.

The spin-down luminosity of J1818 is the highest of all magnetars, and higher than its quiescent X-ray luminosity \citep{hbg+20}. Furthermore, the initial radio spectrum of J1818 was markedly steep (\citealt{2020ATel13560....1M, lsj+20}), but gradually flattened post-outburst \citep{2020ATel13898....1M}. Similar spectral behavior was seen in PSR J1119--1627, a seemingly rotation-powered pulsar which has previously exhibited magnetar-like outbursts (\citealt{akt+16,djw+18}). These properties are shared with rotation-powered pulsars and not radio magnetars, leading previous authors to speculate that J1818 may represent an evolutionary link between pulsars and magnetars \citep{hbg+20}. 

In this paper we detail observations of J1818 taken between 2020 August and 2020 December, ranging in observing frequency from 820 MHz to 35 GHz. Section~\ref{sec:obs} describes the observations and data reduction. The time-averaged properties of the profile and spectrum are described in Section~\ref{sec:prof_spec}, and the single-pulse properties are described in Section~\ref{sec:sps}. We discuss these results and their relation to the magnetar and pulsar population in Section~\ref{sec:disc}, and conclude in Section~\ref{sec:conclusion}.

\input{observations}

\section{Observations and Data Reduction} \label{sec:obs}
We conducted observations of J1818 using the VEGAS backend on the Green Bank Telescope from August to December of 2020. Details of each observation are presented in Table~\ref{tab:obs}. We began each observation with on- and off-source scans of the standard flux calibrator J1445+0958 (OQ172) while firing the pulsed noise diode. 

Except for the 6 and 13-GHz datasets, we also recorded full polarimetric information for each observation. User error prevented the full Stokes parameters from being recorded for the 6-GHz observation, so only total intensity data are available. Furthermore, the noise diodes for the 13-GHz backend were not set up to accurately record full Stokes parameters at the time of our observation, so those data are also summed to total intensity in this study.

As with other magnetars, the spin-down rate of J1818 varied substantially within the first year since its outburst (\citealt{ccc+20,rsl+22}). Given these fluctuations and the sparse cadence of our observations, we did not create a timing model which spanned our entire dataset. We used the \texttt{prepfold} command from the \textsc{presto}\footnote{\url{https://github.com/scottransom/presto}} \citep{presto} software package to fold each observation at the nominal pulsar period of 1.364~s and dispersion measure (DM) of 706 \dmunits $\,$ (\citealt{2020ATel13553....1K,2020ATel13560....1M}), and searched over a range of pulse periods to maximize the signal-to-noise (S/N) of the folded pulse profile. 

Instances of terrestrial radio frequency interference (RFI) were identified and excised using the \texttt{paz} and \texttt{pazi} routines from \textsc{psrchive} \citep{psrchive}. Frequency channels contaminated with continuous, narrowband RFI throughout the entire observation were excised, as well as individual subintegrations dominated by impulsive, broadband RFI signals. After RFI mitigation, each observation was folded and split into 16 subintegrations, and averaged in frequency to create 16 pulse profiles. Using \textsc{presto}, we generated times of arrival (TOAs) by cross-correlating the total integrated profile with  the profile from each sub-integration. The TOAs were passed to \textsc{tempo2}\footnote{\url{https://www.atnf.csiro.au/research/pulsar/tempo2/}} \citep{tempo2}, where we created a basic timing model using the DE405 ephemeris and the known sky coordinates, fitting only for the spin period.

We generated \textsc{psrfits} archives of each observation by folding the search-mode data at the best-fit period using \textsc{dspsr}. The data were polarization- and flux- calibrated using \textsc{psrchive}\footnote{\url{https://psrchive.sourceforge.net}} \citep{psrchive}. We note that for the 6-GHz observation taken on MJD 59092, only total intensity data were recorded. Archives with full single-pulse information were also saved for analysis; the single-pulse analysis is detailed in Section~\ref{sec:sps}.

In order to correct for the Faraday rotation, we applied a fiducial rotation measure (RM) of 1442 \rmunits, as found in previous studies \citep{lsj+20}. To confirm the validity of this RM within our dataset, we fit for the Faraday rotation measure with the \textsc{psrchive} routine \texttt{rmfit} on the 800 MHz observation, which returned a best-fit RM of $1444 \pm 0.2$ \rmunits. Given that \texttt{rmfit} may  underestimate the error bars on these measurements (e.g. \citet{wmg+22}), we determined that the fiducial RM of 1442 \rmunits $\,$ was sufficient.

\section{Profile and Spectral Evolution}
\label{sec:prof_spec}
\subsection{Time-Averaged Pulse Profiles}
\label{sec:prof}

\begin{figure}
    \centering
    \includegraphics[width=\linewidth]{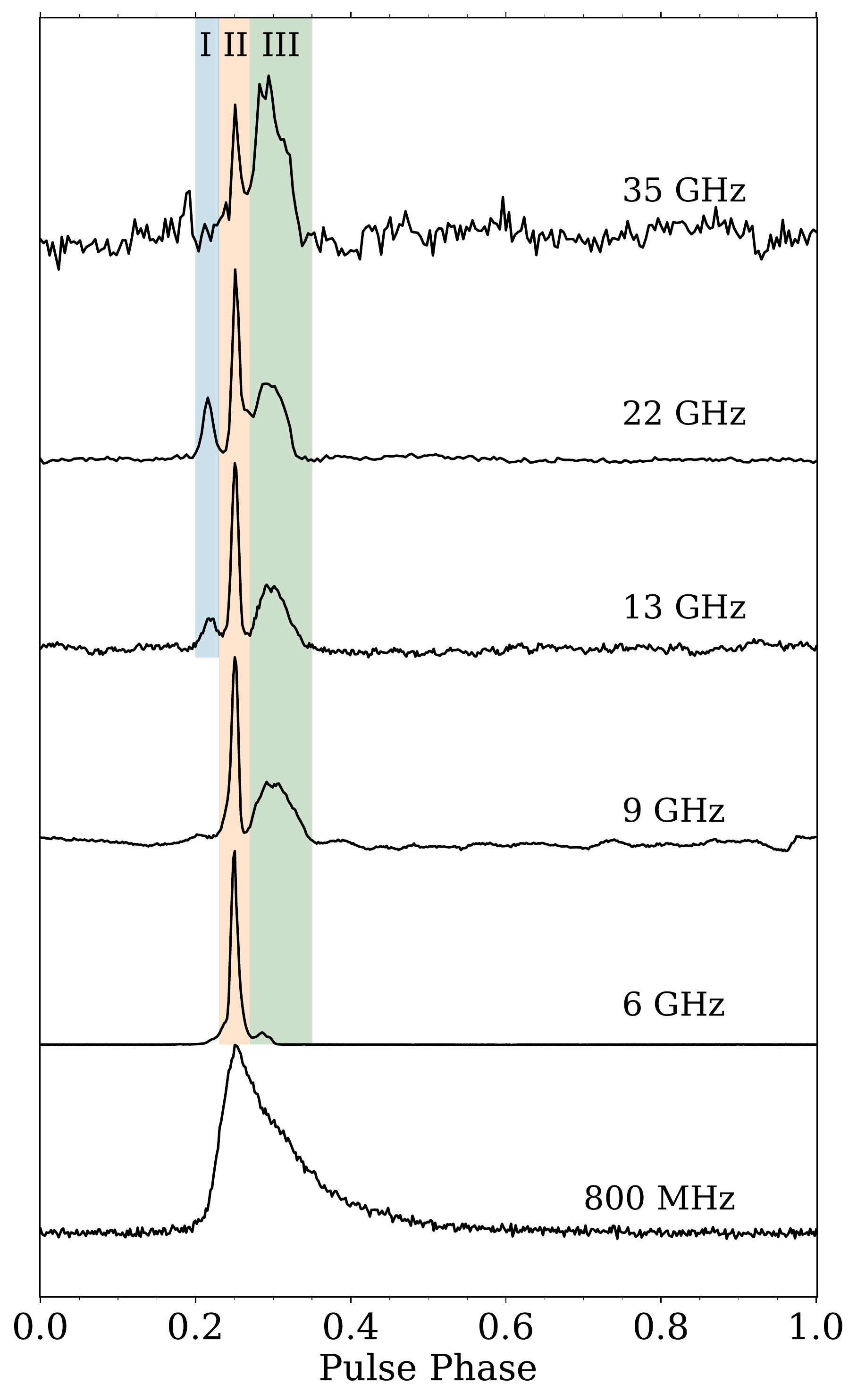}
    \caption{The total intensity profile of J1818 at each observing frequency. Each profile covers one full rotation of the magnetar and is divided into the number of bins which maximizes the profile S/N. For visual clarity, each profile component is highlighted where present: component I in blue, component II in orange, and component III in green.}
    \label{fig:multifreq_profs}
\end{figure}

\input{prof_components}

Figure~\ref{fig:multifreq_profs} displays the time- and frequency-integrated pulse profile of the magnetar at each observing frequency. Above a frequency of 1 GHz, the integrated profiles are made up of three distinct components: a leading component (I), narrow main component (II), and wider trailing component (III). Due to our lack of a phase-connected timing solution, the profiles are all aligned so that the peak of Component II occurs at a rotational phase of 0.25; the 820-MHz profile is aligned so that its peak occurs at the same phase. 

To measure the full-width half-maximum (FWHM) pulse width of each profile component, we used \texttt{scipy.optimize.curve\_fit} to model each component with a single Gaussian function. The resulting pulse widths are reported in Table~\ref{tab:prof_components}.  We also calculate the on-pulse flux of the linearly- and circularly-polarized data, and present the linear polarization fraction (\textit{L/I}) and circular polarization fraction (\textit{C/I}) of each component.

The 820-MHz profile is dominated by scatter broadening. The NE2001 electron-density model predicts a scattering timescale\footnote{Calculated using \url{https://apps.datacentral.org.au/pygedm/}} of  61 ms \citealt{ne2001}). An alternative approach is to use the empirical relation between DM and scattering timescale; this results in a prediction of 72 ms at that frequency \citep{ymw17}. We estimated the scattering timescale by fitting to the 820-MHz integrated profile a Gaussian function convolved with a one-sided exponential broadening function, and scaled to 1 GHz assuming a frequency dependence of $\tau_s \sim \nu^{-4.4}$. The best-fit scattering timescale of $46 \pm 4$ ms is in agreement with the previously measured $\tau_s = 42_{-3}^{+9}$ ms \citep{lsj+20} and $44 \pm 3$ ms \citep{ccc+20}.  Above frequencies of 6 GHz, the scattering timescale is predicted to be less than 10 ns and the effects of scattering can be ignored. 

The polarization position angle (PA) of the linear polarization, $\Psi$, is shown for each observation in the top panel of Figure~\ref{fig:freqphase}. Variations in the PA with pulse phase can be explained with the rotating vector model (RVM); the RVM models the pulsar magnetic field as a simple dipole, in which case the PA `sweep' across the radio profile is caused by the changing angle between the magnetic axis and our line of sight \citep{rvm_rc}. In this simple case, the RVM predicts a flat, S-shaped sweep in PA across the pulse profile. 

The PA swing at 820 MHz is flatter than at higher frequencies, while still retaining the expected S-shaped curve. The relatively flat slope is likely caused by interstellar scattering \citep{kj08}, and agrees with earlier low-frequency detections of J1818 (\citealt{ccc+20,lsj+20}).
The PA swing of the 22 GHz observation is much steeper, and inverted with respect to the other observations, but still retains the characteristic S-shaped sweep. Similar changes in the polarization properties of J1818 were seen at 2.6 GHz by \citet{ljs+21}, in which case the PA exhibited a steeper, reversed PA sweep on a single epoch of observation before returning to a positive slope.

The spectral properties of the individual components are further discussed in Section~\ref{sec:phase-spec}, and single-pulse properties in Section~\ref{sec:sps}.

\subsection{Flux Density and Spectral Indices} \label{sec:spectrum}

\begin{figure*}
    \centering
    \includegraphics[width=\textwidth]{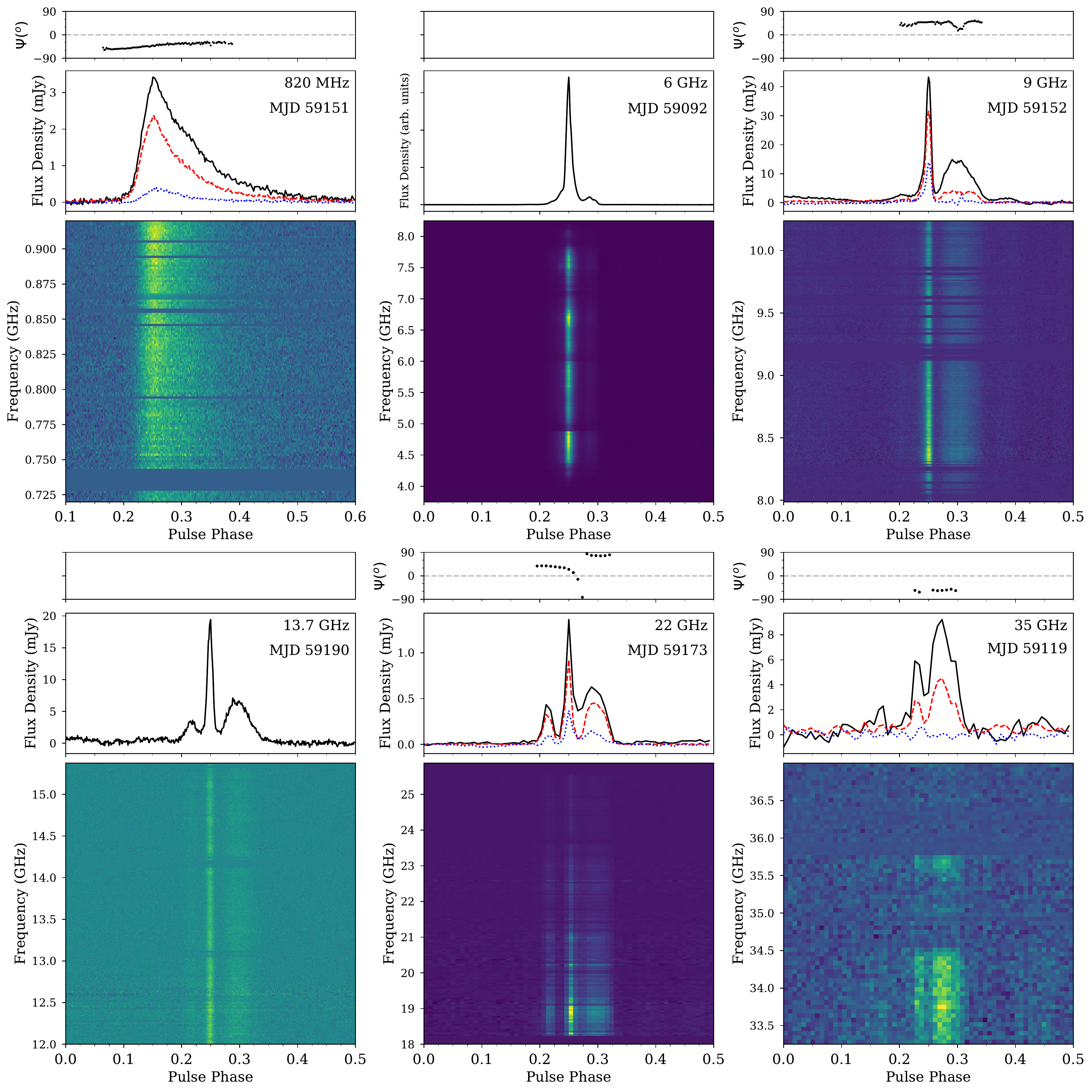}
    \caption{For each observation of J1818, we show the polarization position angle sweep (top panel), calibrated polarization profile with total intensity in black, linear polarization in red, and circular polarization in blue (middle panel), and frequency-phase plot (bottom panel). Polarization information is not available for the 6- and 13-GHz data.}
    \label{fig:freqphase}
\end{figure*}

\begin{figure*}
    \centering
    \includegraphics[width=\textwidth]{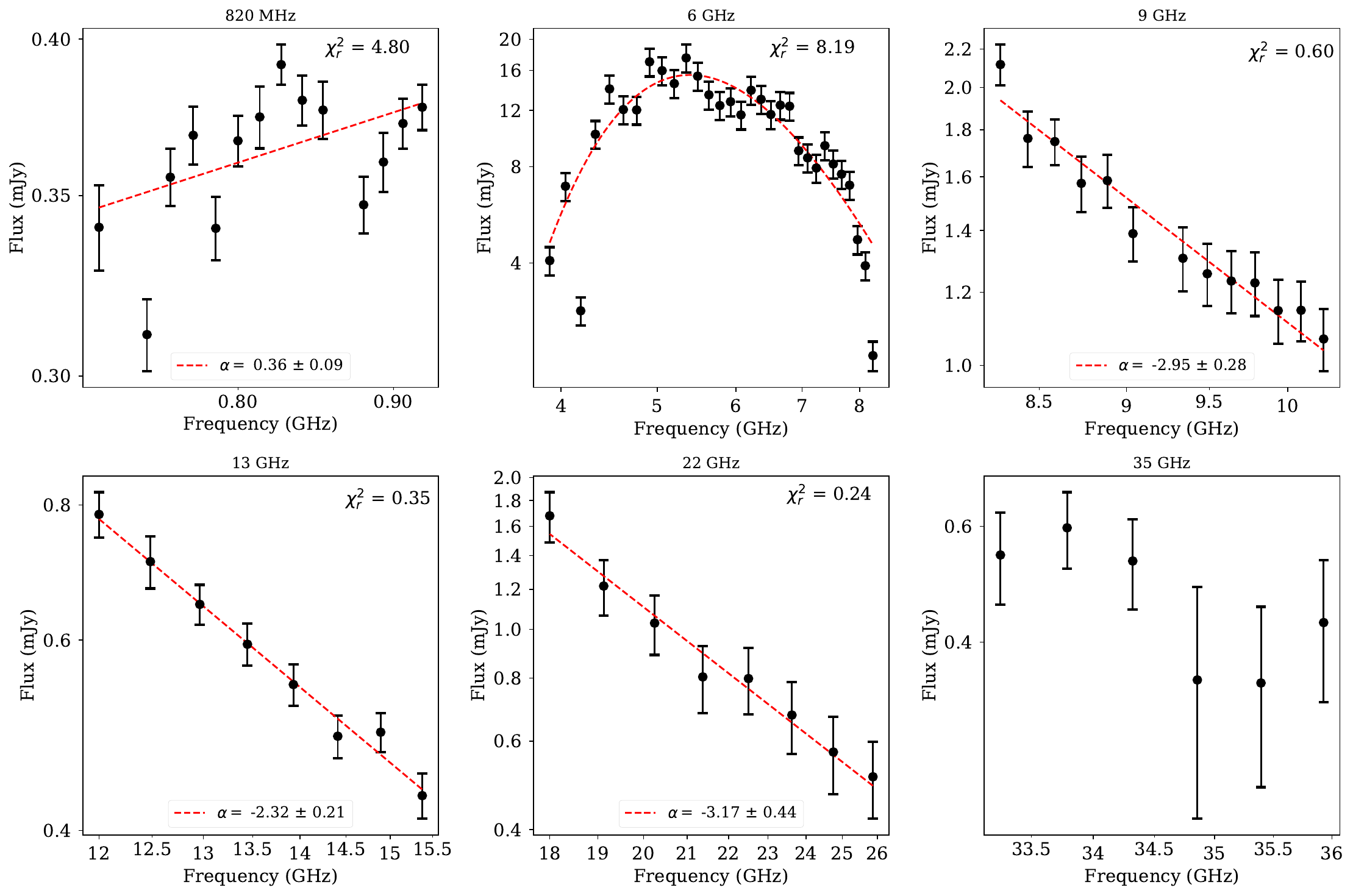}
    \caption{The flux measurements in each subband and their corresponding errors are shown in black. All of the y-axes are in log scale. The best-fit power law is overlaid in red, with the associated spectral index and 1-$\sigma$ error at the bottom of each panel. The 6-GHz data are fit with a combined power-law and free-free absorption spectral model (see text). No spectral index was obtained when fitting the 35-GHz data. The reduced chi-squared values associated with each fit are given in the top right corner of each sub-panel.}
    \label{fig:spectra_grid}
\end{figure*}

Figure~\ref{fig:freqphase} shows the time-integrated frequency/phase plots, polarization profiles, and linear polarization position angles for each observation in which the magnetar was detected. We see no evidence of astrophysical scintillation in our dataset. Below 22 GHz, the scintillation bandwidth predicted by the NE2001 model is smaller than the bandwidth of a single frequency channel. The scintillation bandwidth predicted by the NE2001 model is 2.4 MHz in the 22-GHz observing band, and 19 MHz in the 35-GHz observing band. Therefore, even though just resolvable, it would be difficult to detect scintillation with sufficient signal-to-noise at these frequencies. 

In order to measure the flux density of the source at each observing frequency, each observation was time-integrated and split into at most 16 subbands to produce a one-dimensional pulse profile for each subband.

The number of subbands was chosen such that the measured flux density was greater than its corresponding error in each subband; higher-significance flux measurements would have necessitated using too few subbands to get good fits. The mean flux density $S$ in each subband is given by 

\begin{equation}
S = \frac{1}{N_b} \sum_{i=0}^{N_b} S_i,
\end{equation}

\noindent
where $N_b$ is the number of bins in the profile, and $S_i$ is the flux density in bin $i$. We first subtract the off-pulse mean from each profile, so this method is equivalent to summing over only the on-pulse phase bins. The on-pulse phase window is defined using the phase ranges 0.2 to 0.35, except for the 820-MHz profile, where the on-pulse region is extended to 0.6 to account for scatter broadening. The off-pulse region is defined as the regions not covered by the on-pulse region: phase ranges 0 to 0.2, and 0.35 to 1 (0.6 to 1 for the 820-MHz profile). The corresponding error on each flux density measurement is calculated from the off-pulse root mean square (RMS) divided by the square root of the number of phase bins.

We also summed each observation into a single frequency channel and measured the mean flux of J1818 at that observing frequency; the results are given in Table~\ref{tab:spidx}.

\input{spidx} 

The 6 GHz observation did not have available flux calibration, so we used the radiometer equation to scale each profile into Jansky units using a scaling factor:

\begin{equation}
    \frac{\mathrm{T_{sys}} }{\sigma \mathrm{G} \sqrt{\mathrm{n_p} t_{obs}\Delta\nu}}
\end{equation}

The system temperature $\mathrm{T_{sys}} = \mathrm{T_{rec}} + \mathrm{T_{sky}}$, where the receiver temperature $\mathrm{T_{rec}}=23$ K (``Observing with the GBT'', version 4.0)\footnote{\url{https://greenbankobservatory.org/portal/gbt/observing/}} and the sky temperature $\mathrm{T_{sky}}$ was calculated at the central frequency of each subband using the \texttt{pyGDSM} Python interface\footnote{\url{https://github.com/telegraphic/pygdsm}} for the \citet{ztd+17} global diffuse sky model. The gain of the GBT, $G$, is 2 K/Jy (``The Performance of the GBT'')\footnote{\url{https://www.gb.nrao.edu/GBT/Performance/PlaningObservations.htm}}, and the on-source time, $t_{obs}$, is taken from Table~\ref{tab:obs}. The bandwidth ($\Delta \nu$) and off-pulse standard deviation ($\sigma$) are calculated separately for each subband. The number of recorded polarizations, $n_p$, is two for the full Stokes observations. We add in quadrature a 10\% uncertainty on each flux measurement due to the systematic errors involved in using the radiometer equation, due to uncertainties and time dependence in the gain and system temperature.

Flux measurements of the calibrated 22-GHz observation yielded mean fluxes on the order of 50 $\mu$Jy, at odds with the high S/N of its integrated profile and the mean flux of J1818 at all other frequencies. We used the radiometer equation as a validity check, assuming a system temperature of 40 K and using the same method described above to estimate the fluxes and their corresponding errors. The radiometer equation returns mean fluxes in the range of 0.5 to 1.7 mJy, and given that these measurements made more physical sense, we determine that there was an error in the flux calibration, and use the radiometer fluxes from the 22-GHz observation for our analyses. 

The evolution of flux with frequency of pulsars is usually described by a power law relationship, with a spectral index $\alpha$ where $S \sim \nu^{\alpha}$. Since the majority of our spectra in Figure~\ref{fig:spectra_grid} appeared to follow power-law relationships, we fit power-law spectra to those flux measurements using the non-linear least squares fitting method provided by \texttt{scipy.optimize.curve\_fit}. The flux measurements and their associated spectral fits are shown in Figure~\ref{fig:spectra_grid}, and Table~\ref{tab:spidx} reports the measured spectral indices for each observation, with 1-$\sigma$ error bars. The 35-GHz signal-to-noise ratios in individual subbands were too low to accurately constrain a spectral index. 

The 6-GHz spectrum is not defined by a single power law. Given the clear spectral peak within the observing band, and the negative spectral indices at all higher frequencies, we see evidence of a gigahertz-peaked spectrum (GPS) as seen in other magnetars (discussed further in Section~\ref{sec:disc}). To model the spectral behavior, we used a simple power law combined with a free-free absorption model as described in \citet{kbl+21}. The evolution of flux with frequency is modeled as 

\begin{equation}
    S(\nu) = A \left(\frac{\nu}{\nu_0}\right)^{\alpha} e^{-B \,\nu^{-2.1}}
\end{equation}

where $A$ is the intrinsic flux of the pulsar at a characteristic frequency $\nu_0$, $\alpha$ is the power-law spectral index, and $B$ is a frequency-independent parameter. 

The observed radio spectrum peaks at a frequency of 5.35 GHz. The best-fit model, obtained through least-squares fitting, peaks at a frequency of $5.4 \pm 0.6$ GHz. The GPS spectral model does not directly fit for a peak frequency, so to estimate an error, we used the 1-$\sigma$ errors on the spectral fit to find the peak frequency of the models with $\pm 1 \sigma$ on the parameters $A$, $\alpha$, and $B$. 

The mean flux of J1818 at each frequency is shown in the top panel of Figure~\ref{fig:phase_spectra}. The light gray bars denote the frequency ranges of the 9- and 22- GHz datasets, to distinguish them from the other observations. We note that these observations are spread out over the course of roughly 100 days, and given the known flux variability of magnetars over time as well as the long-term emission mode changes reported by \citet{rsl+22} and \citet{fbr+24}, these results cannot be used to study the instantaneous broadband spectrum of J1818 nor the time variability of its spectrum. 

\subsection{Phase-resolved Spectroscopy} \label{sec:phase-spec}

\begin{figure*}
    \centering
    \includegraphics[scale=0.75]{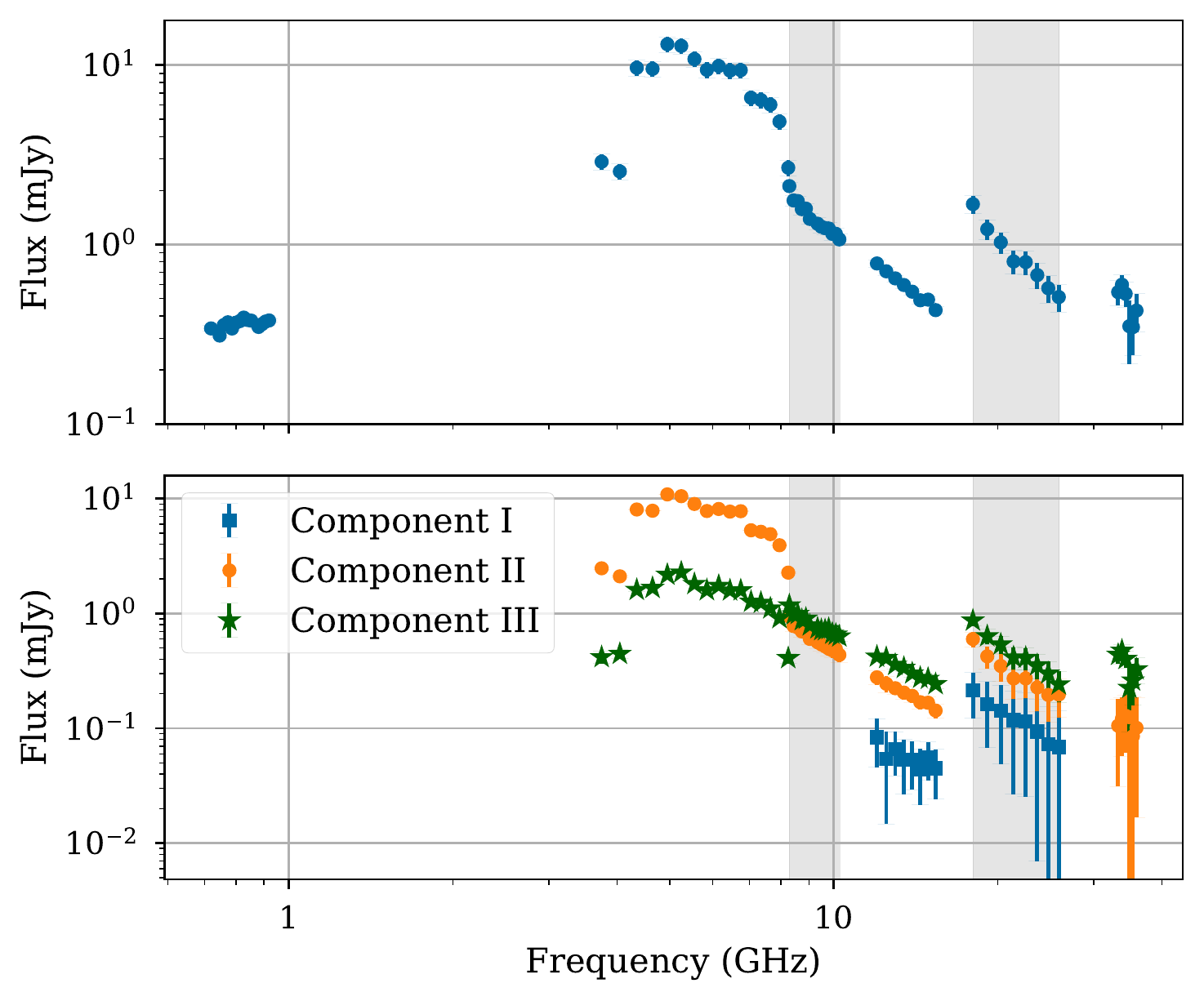}
    \caption{Top: phase-averaged mean flux of J1818 as a function of observing frequency. Bottom: the phase-averaged mean flux of each individual pulse component as a function of observing frequency. The frequency ranges of the 9- and 22-GHz observations are indicated with vertical gray bars to separate them from the other observations.}
    \label{fig:phase_spectra}
\end{figure*}

In order to better understand the emission properties of the individual profile components, we also measured the time-averaged spectral index of each component. We defined the on-pulse region of each individual component using the highlighted phase ranges in Figure~\ref{fig:multifreq_profs}, and calculated mean on-pulse fluxes for each component by using only these relevant parts of the profile. We used these on-pulse fluxes to calculate the spectral indices of the individual profile components at each frequency. The mean fluxes of each component are shown in the bottom panel of Figure~\ref{fig:phase_spectra}, and the resulting fitted spectral indices are provided in Table~\ref{tab:prof_components}.


\section{Single-Pulse Analysis} \label{sec:sps}

\begin{figure}
    \centering
    \includegraphics[scale=0.5]{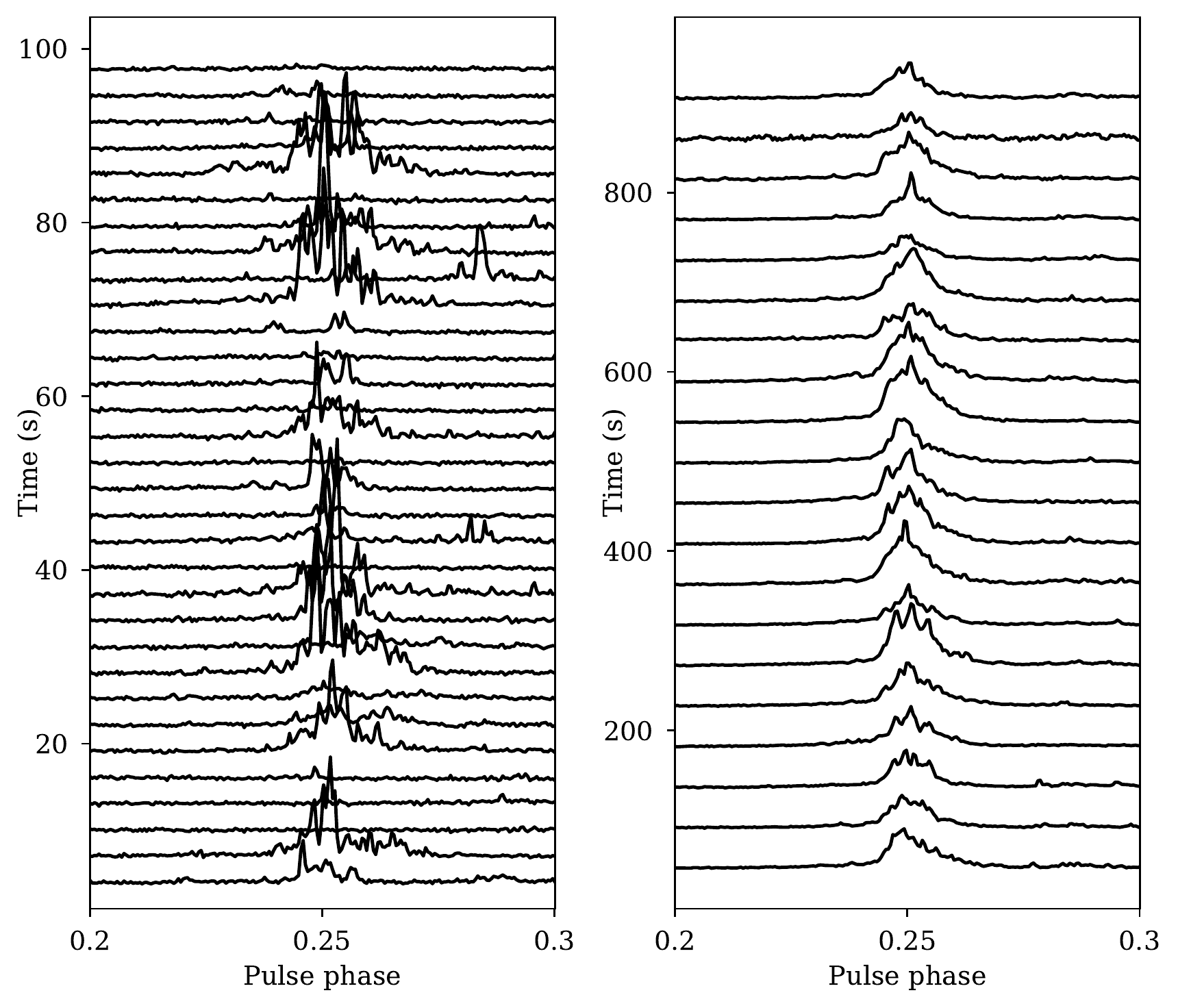}
    \caption{Stacked pulse profiles from the 6-GHz observation of J1818. The pulses are centered at a pulse phase of 0.25 and the horizontal axis covers one tenth of the magnetar's rotation. Left: 32 consecutive pulses stacked atop one another. Each profile represents one rotation of the magnetar. Right: the change in the integrated profile at 6 GHz over time. Each profile represents 32 rotations of the magnetar, equivalent to 43.6 seconds.}
    \label{fig:Cband_pulsestacks}
\end{figure}

\begin{figure*}
    \centering
    \includegraphics[width=\linewidth]{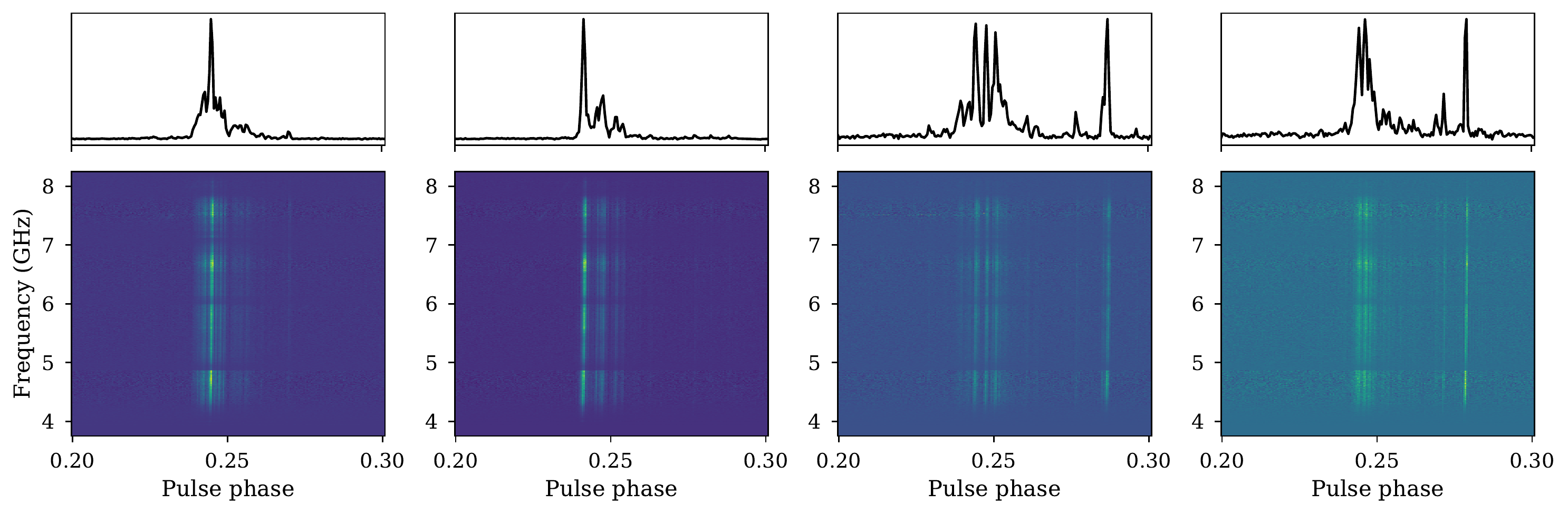}
    \caption{Selected single pulses from the 6-GHz observation. Top panel: the intensity of each single pulse as a function of time. Bottom panel: intensity of the pulse as a function of frequency and time. We show 10\% of the full pulse period.}
    \label{fig:C_sps}
\end{figure*}

\begin{figure*}
    \centering
    \includegraphics[width=\textwidth]{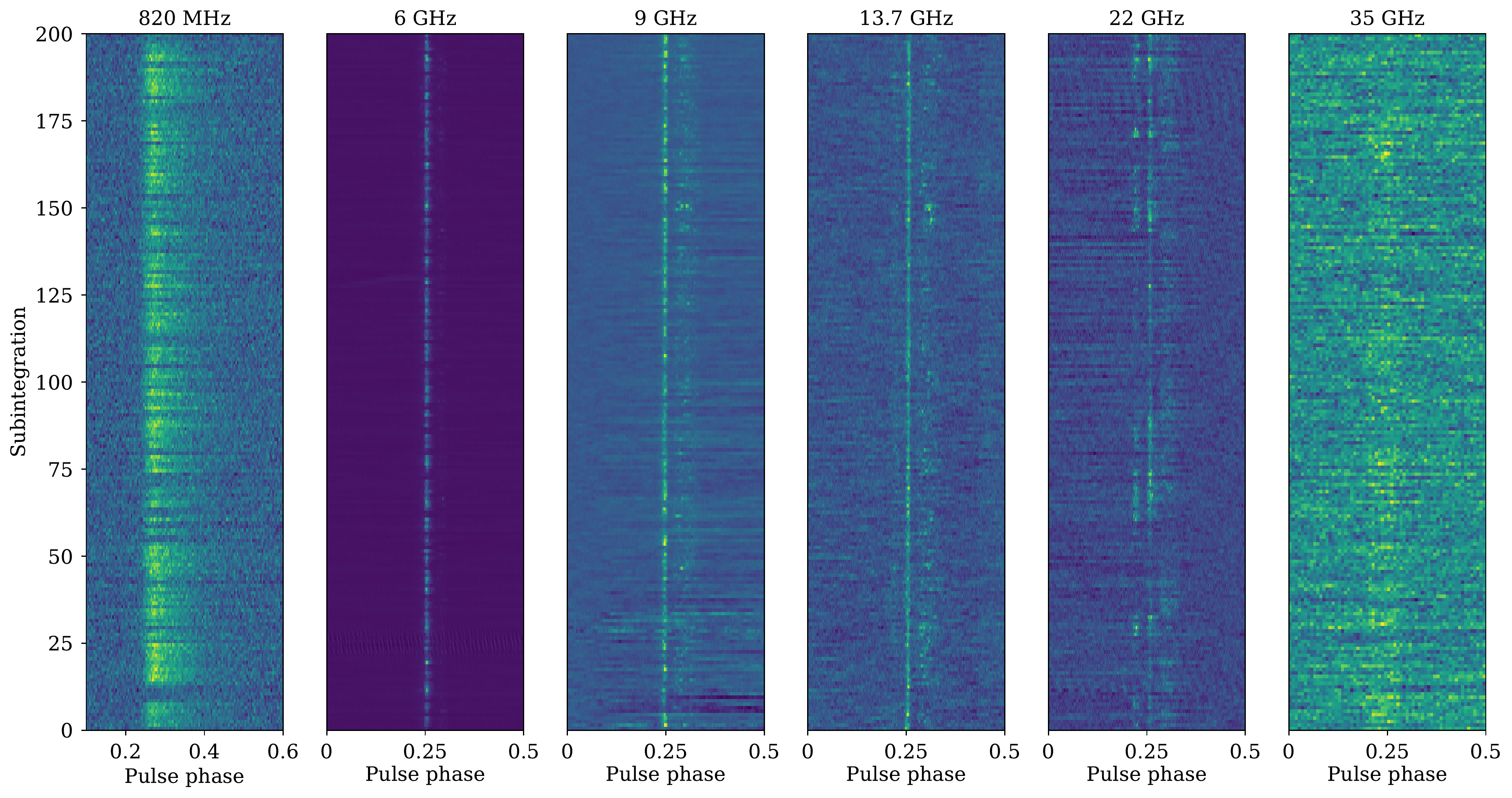}
    \caption{Subintegration pulse stacks of each observation of J1818 analyzed in this work. Each subintegration has a duration of roughly 9.55 seconds, equivalent to seven rotations of the magnetar.}
    \label{fig:timephase}
\end{figure*}

\input{spstats_with_flux}

The single pulses of J1818 demonstrate significant variability in their flux and morphology over time, similar to other radio magnetars. For instance, Figure \ref{fig:Cband_pulsestacks} displays 32 consecutive single pulses at 6 GHz, contrasted against the relatively stable pulse profile over the 3.4-hour long observation. A selection of single pulses at 6 GHz and their waterfall plots are also shown in Figure~\ref{fig:C_sps}. Time-phase plots with 9.55-second (i.e. seven pulses) subintegrations are shown in Figure~\ref{fig:timephase}.

\subsection{Pulse Energies}
\label{sec:pulse_energies}
\begin{figure}
    \centering
    \includegraphics[width=0.8\linewidth]{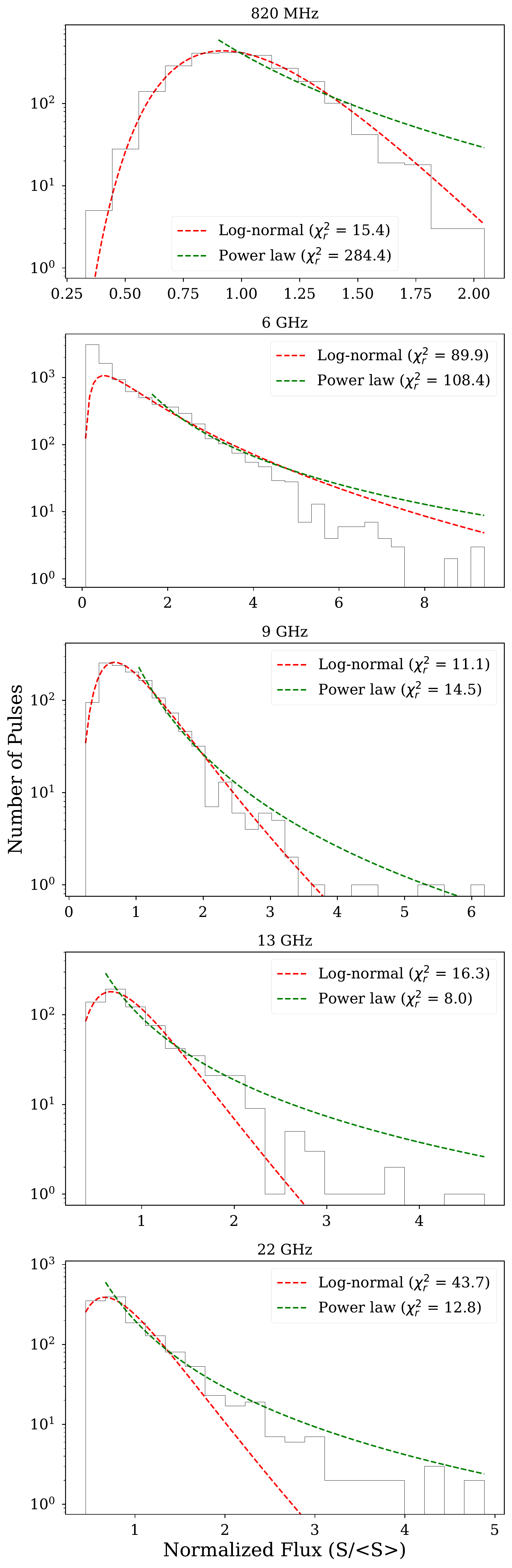}
    \caption{The distributions of phase-averaged peak flux density are shown in black for each observation. The pulse fluxes are normalized by the mean of the peak flux densities of all pulses from that epoch. Overlaid in red are the log-normal (red dashed line) and power-law tail (green dashed line) fits to each distribution. The reduced $\chi^2$ values for each fit are given in the legend. There were not enough high-significance pulses at 35 GHz to fit a distribution.}
    \label{fig:pulse_energy_fits}
\end{figure}

We measured the peak flux of each pulse using the peak amplitude of the on-pulse region. The significance of each pulse was calculated as the peak amplitude divided by the off-pulse standard deviation. Only pulses with significance $\geq$ 4 were considered. The fluxes of each pulse were normalized by the mean peak flux of all pulses on that epoch. 

We fit a log-normal and power-law function to each distribution of pulse fluxes, except for the 35-GHz distribution due to the low number of high-significance single pulses. We fit the power-law distributions to energy bins past the peak of the distribution. The peak flux distributions and their corresponding fits are shown in Figure~\ref{fig:pulse_energy_fits}. At low observing frequencies, the log-normal fits are preferred, with no evidence of a power-law tail. Above 9 GHz, the best-fitting log-normal distribution underestimates the number of high-flux pulses, and the power-law tail fits are preferred, though we note that the power-law fits still overestimate the number of high-flux pulses, especially at 6 and 13 GHz.

\begin{figure*}
    \centering
    \includegraphics[width=\textwidth]{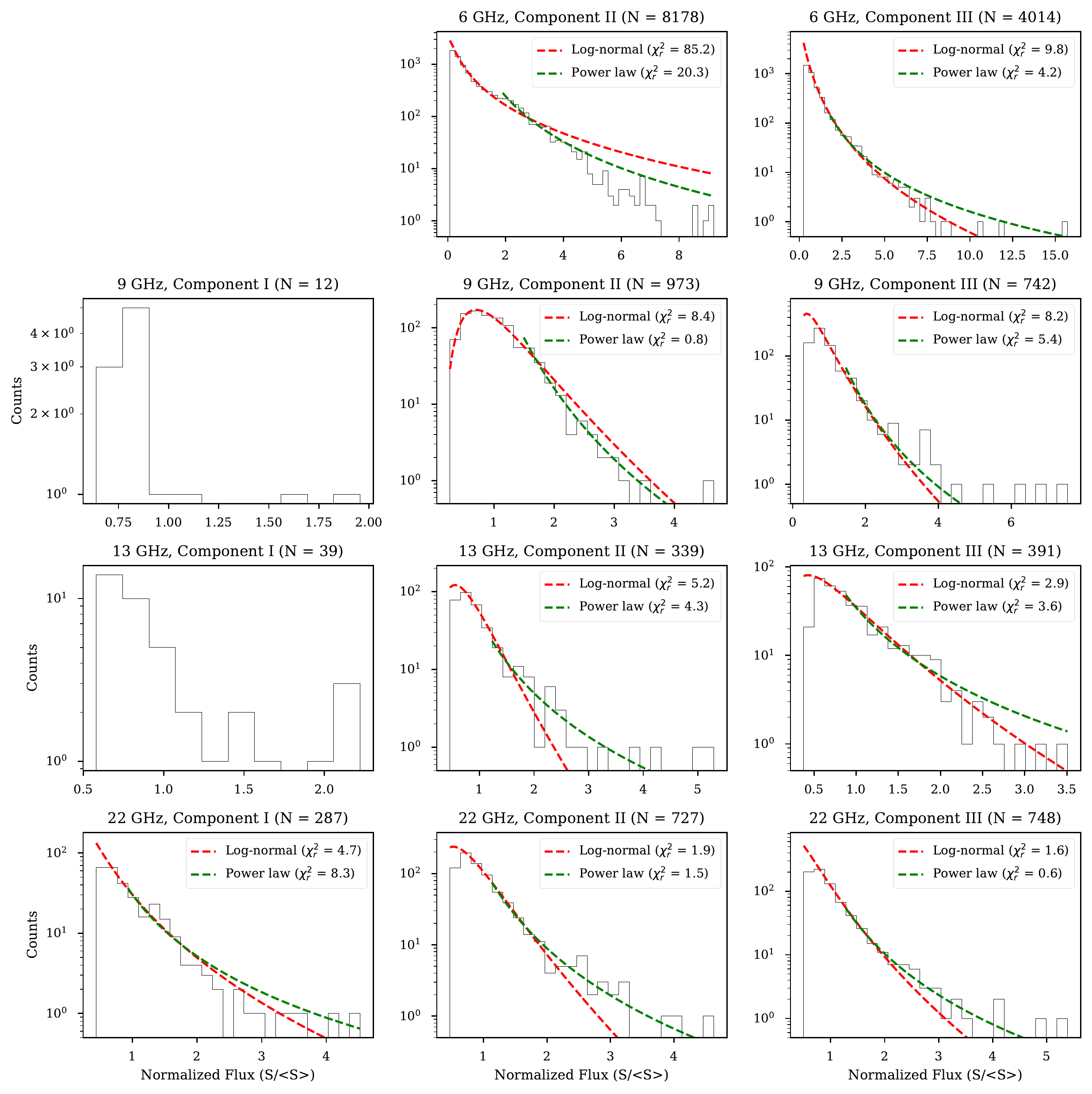}
    \caption{The normalized peak flux distributions of each individual component from 6 to 22 GHz. The x-axis is normalized by the mean peak flux of that individual profile component. Overlaid in red are the log-normal (red dashed line) and power-law tail (green dashed line) fits to each distribution. The reduced $\chi^2$ values for each fit are given in the legend. Fits to the Component I distributions at 9 and 13 GHz failed due to the small number of pulses.}
    \label{fig:pulse_energy_components}
\end{figure*}

Since each single pulse is represented by its peak flux across the on-pulse region, Figure~\ref{fig:pulse_energy_fits} only considers the brightest component within each rotation. We measured the peak flux for each profile component across each rotation of the magnetar, and normalized them by the mean peak flux of that component on that epoch. Figure~\ref{fig:pulse_energy_components} shows the distribution of pulse energies, separated by pulse component. The only pulses with energies $>10$ times the average energy (a typical definition for giant pulses) come from Component III at 6 GHz, for which the power law fit is preferred. 

\subsection{Flux Modulation}
\begin{figure*}
    \centering
    \includegraphics[scale=0.5]{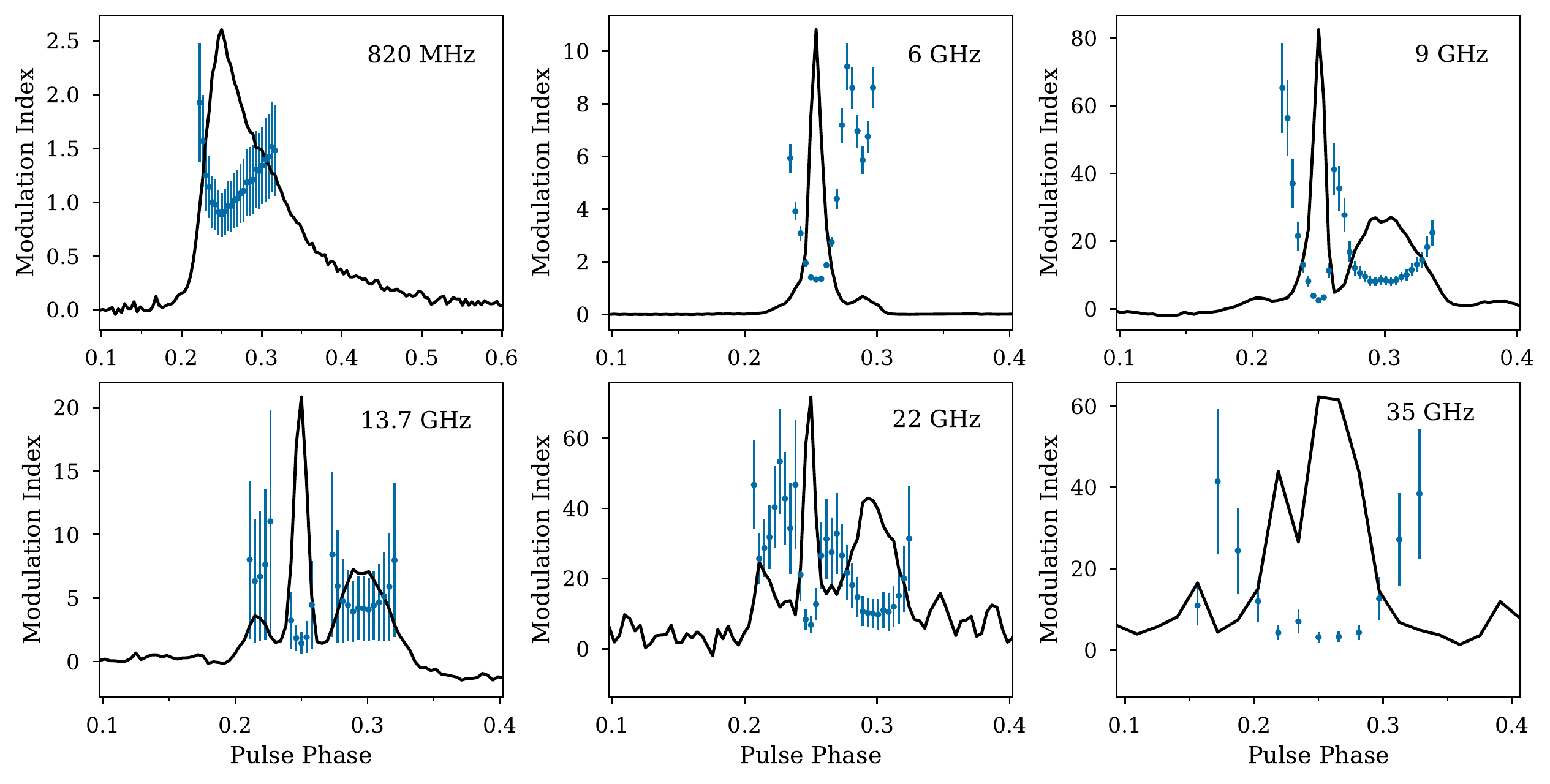}
    \caption{The modulation indices of J1818 as a function of pulse phase at each frequency are shown in blue. The normalized integrated pulse profiles are overlaid in black. Modulation indices are not plotted for phase bins dominated by noise.}
    \label{fig:modindex}
\end{figure*}

\begin{figure*}
    \centering
    \includegraphics[scale=0.45]{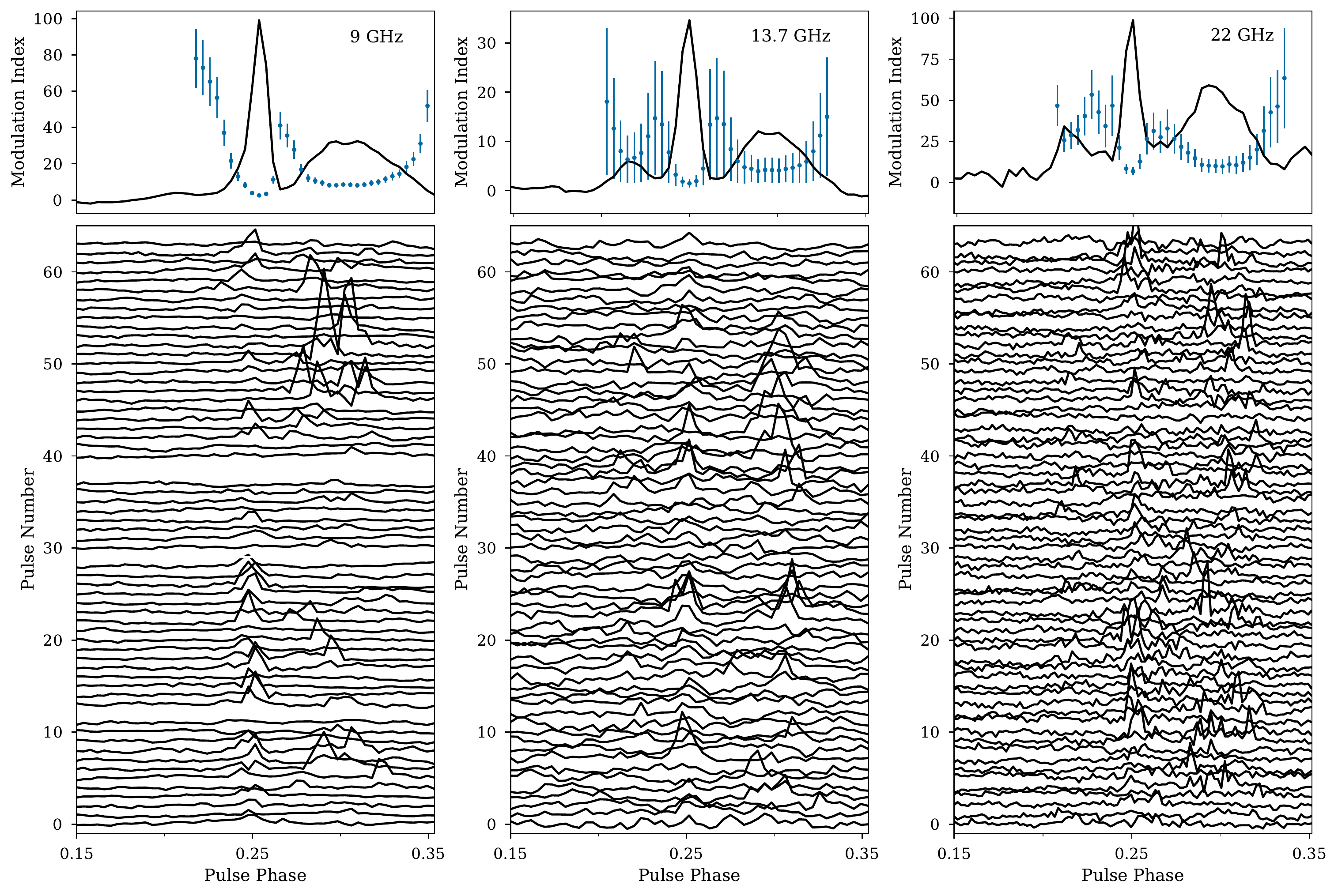}
    \caption{The modulation indices of J1818 at 9, 13, and 22 GHz, with pulse stacks. Top panel: modulation indices are plotted in blue, normalized integrated pulse profiles are overlaid in black. Bottom panel: 64 consecutive single pulses at each frequency. Some pulses have been excised due to RFI corruption.}
    \label{fig:zoomed_modindex}
\end{figure*}

To quantify the temporal variability of each pulse profile component, we calculated the modulation index of each phase bin. We calculate the integrated profile, variance, and the modulation index of each phase bin following the formulation laid out in \citet{wes+06}, which we reproduce here. The value of the mean folded profile, $\mu_i$ at bin $i$, is 

\begin{equation}
    \mu_i = \frac{1}{N} \sum_{j=0}^{N} S_{ij}
\end{equation}

where $N$ is the total number of pulses averaged over, and $S_{ij}$ is the signal intensity in pulse bin $i$ in pulse $j$. We calculate the variance of intensities within each pulse bin,  

\begin{equation}
    \sigma_i^2 = \frac{1}{N} \sum_{j=0}^{N} (S_{ij} - \mu_i)^2
\end{equation}

and the modulation index of pulse bin $i$ is then 

\begin{equation}
    m_i = \frac{\sigma_i}{\mu_i}
\end{equation}

Figure~\ref{fig:modindex} displays the modulation indices across the on-pulse window of J1818 at each frequency, and Figure~\ref{fig:zoomed_modindex} displays the modulation indices at 9, 13, and 22 GHz along with pulse stacks. 

At each frequency, the modulation index of component II is significantly smaller than those of components I and III, where detected. As seen in the pulse stacks in the bottom panels of Figure~\ref{fig:zoomed_modindex}, the modulation indices suggest that component II would have a relatively consistent intensity over time, and the side components display more infrequent and variable emission, comparable to the narrow, spiky subpulses typical to magnetar emission.

\subsection{Polarization}
\begin{figure}
    \centering
    \includegraphics[width=\linewidth]{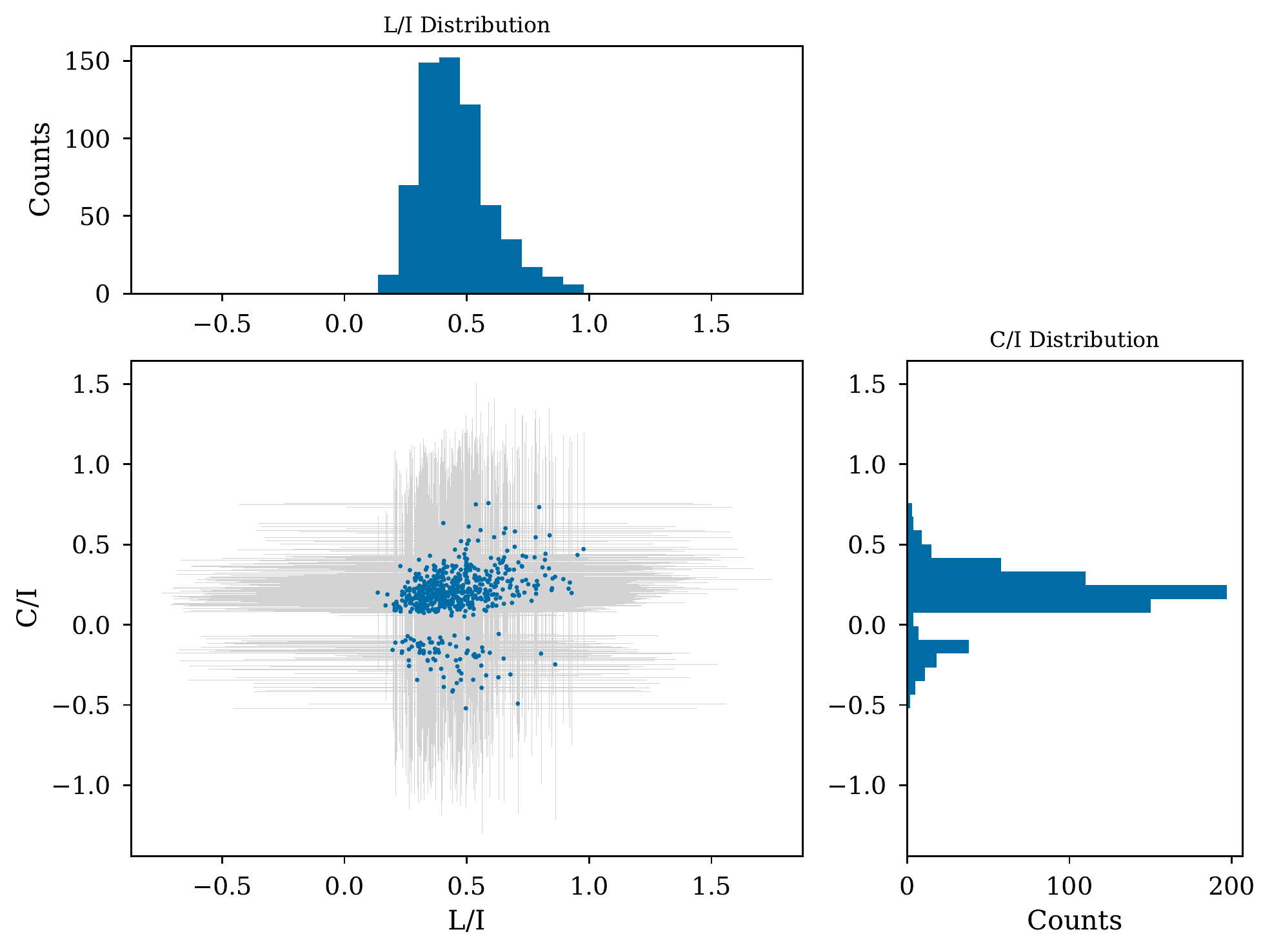}
    \caption{The distribution of linear polarization fraction (L/I) and circular polarization fraction (C/I) for each single pulse at 820 MHz. 1-$\sigma$ error bars on each point are denoted in light gray. The mean values of the polarization fractions are shown in the histograms with black dashed lines.}
    \label{fig:polfracs_820}
\end{figure}

\begin{figure}
    \centering
    \includegraphics[width=\linewidth]{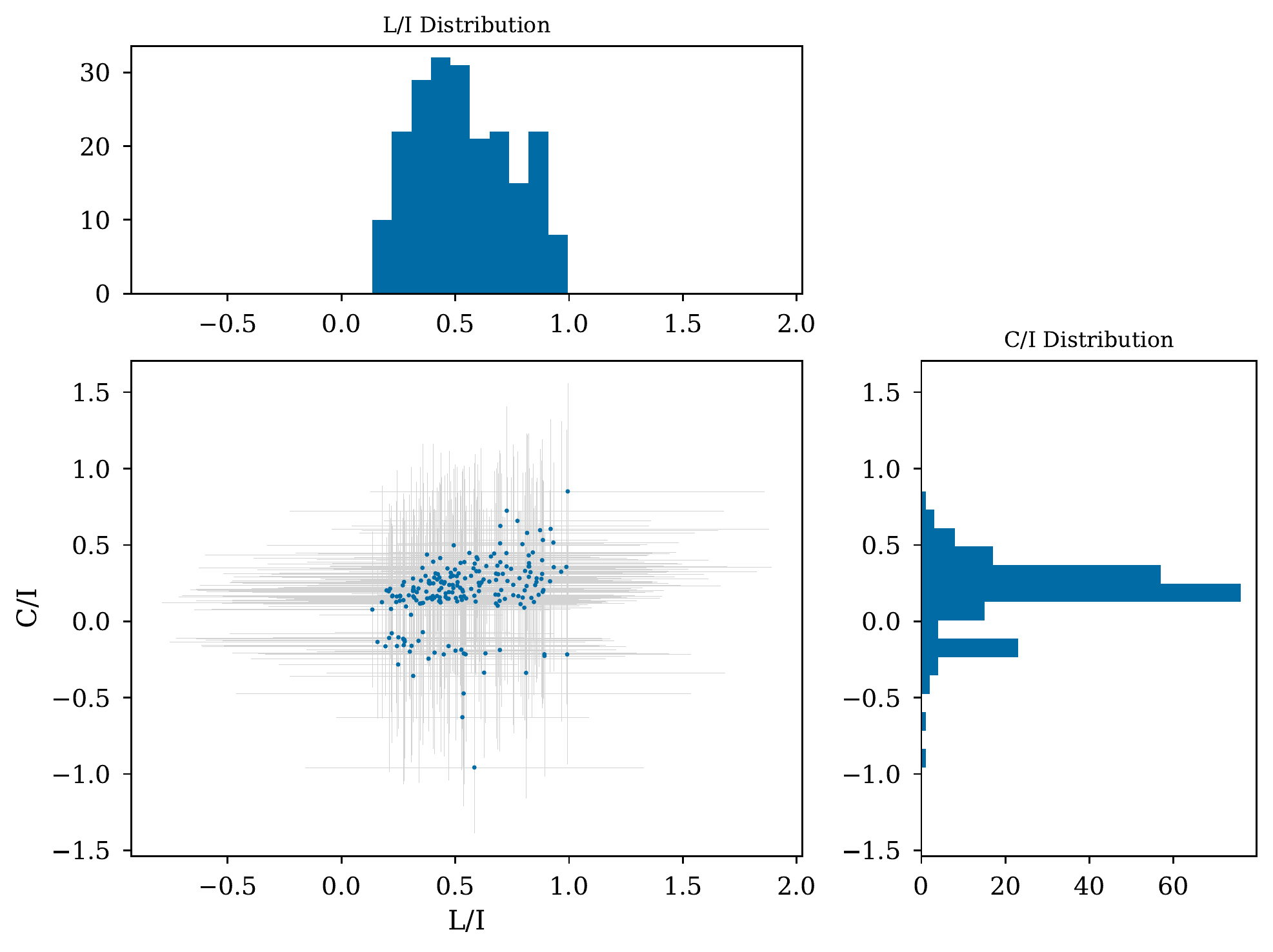}
    \caption{The distribution of linear polarization fraction (L/I) and circular polarization fraction (C/I) for each single pulse at 9 GHz. 1-$\sigma$ error bars on each point are denoted in light gray. The mean values of the polarization fractions are shown in the histograms with black dashed lines.}
    \label{fig:polfracs_9}
\end{figure}

\begin{figure*}
    \centering
    \includegraphics[scale=0.6]{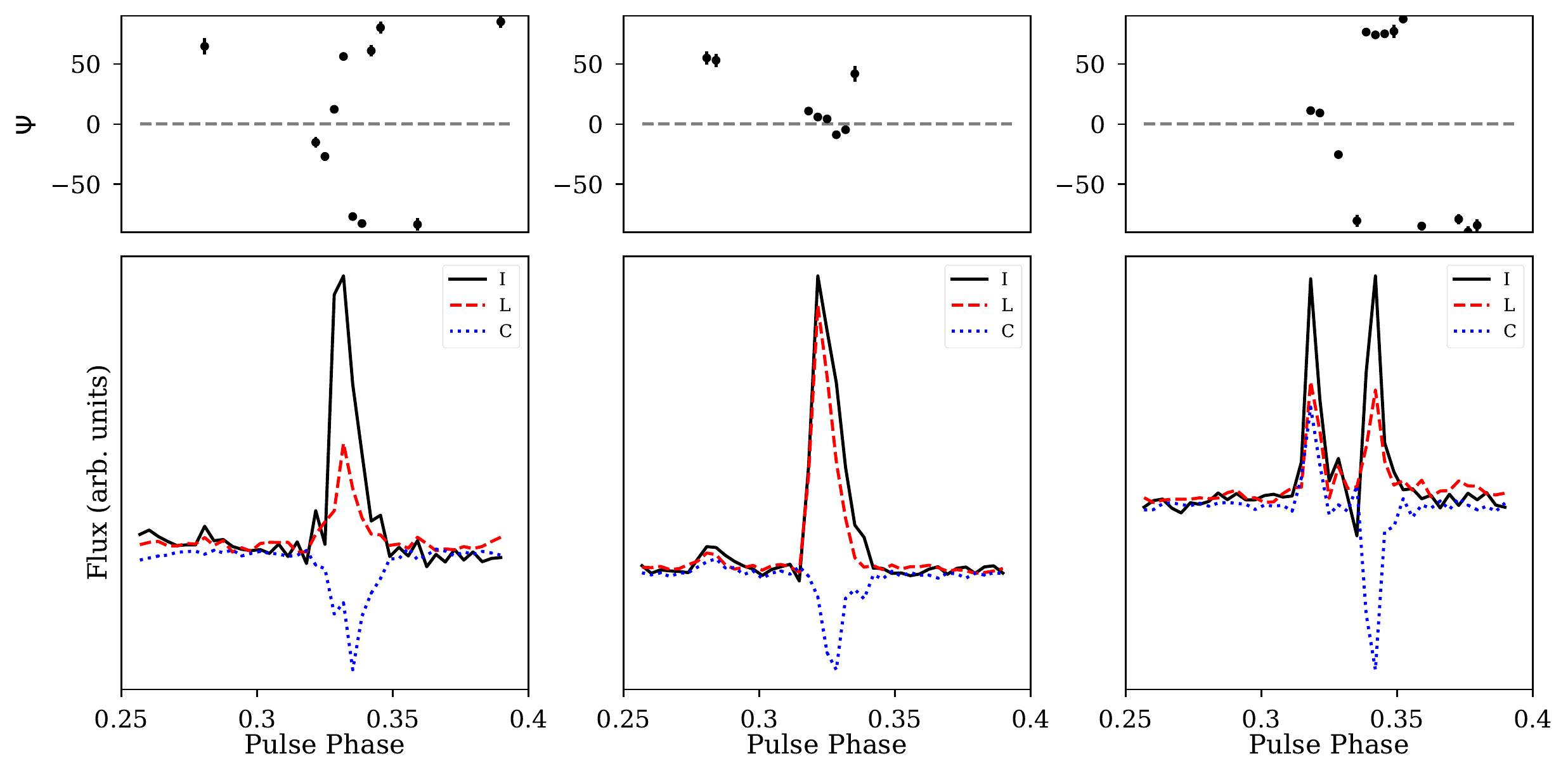}
    \caption{Example single pulses with reversed handedness of circular polarization. Top panels: the polarization position angle, with a gray horizontal dashed line at zero degrees. Bottom panels: the total intensity profile of the single pulse in solid black, linear polarization as a dashed red line, and circular polarization as a dotted blue line.}
    \label{fig:neg_CP}
\end{figure*}

We examined the degree of linear and circular polarization within each single pulse by measuring the polarized flux and comparing it with the flux of the total-intensity pulse. The errors on the linearly and circularly polarized fluxes are calculated as the off-pulse RMS of the respective polarized pulse profile (i.e. the RMS of the linearly polarized profile is the error on the linearly polarized flux).  To examine the distribution of polarization fractions, we first excised pulses from our sample which had insufficient S/N to adequately measure the flux.


The resulting distributions of single pulse polarization fractions (linear, \textit{L/I}, and circular, \textit{C/I}) at 820 MHz and 9 GHz are respectively shown in Figures~\ref{fig:polfracs_820} and~\ref{fig:polfracs_9}. The mean values ($\mu$) of the single pulse polarization fractions are denoted in the histograms. In comparison to the polarization fractions of the time-integrated profiles given in Table~\ref{tab:prof_components}, the pulses at 820 MHz have a lower average \textit{L/I} and higher average \textit{C/I}. At 9 GHz, the \textit{L/I} distribution is wider and there are more highly linearly polarized pulses, but the \textit{C/I} distributions are similar at both frequencies. 


At both frequencies, a portion of the single pulses have the sign of the circular polarization reversed. Pulse-to-pulse changes in the handedness of the circular polarization have been observed before in radio magnetars alongside structures in the PA which deviate significantly from the RVM \citep{ksj+07,dwg+24,ljl+24}.  Similar behavior was also observed in a long-period radio transient with similar emission properties to FRBs and magnetars \citep{mmh+25}. We show some of the 9-GHz single pulses and their PA swings in Figure~\ref{fig:neg_CP}. 

\subsection{Searches for Periodic Substructure}

Across all frequencies, the single pulses from J1818  are comprised of multiple emission components of varying widths and strengths. Even within a single `component' region as described in Section~\ref{sec:prof}, there may be several bright, narrow peaks (see e.g. Figures~\ref{fig:Cband_pulsestacks}, \ref{fig:C_sps}, and \ref{fig:zoomed_modindex}). 

The complexity and pulse-to-pulse shape variation discouraged us from attempting a simple FWHM width fit to each pulse component. Instead, we employed an autocorrelation function (ACF) to uncover any underlying quasi-periodicities within the on-pulse emission region of each pulse. Quasi-periodic structure within individual pulses has been suggested as a common feature between all radio-emitting neutron stars, including rotation-powered pulsars and magnetars \citep{kld+24}. Though the exact value of the sub-periodicities can vary from pulse to pulse and between epochs, \citet{kld+24} found characteristic quasi-periodicities of 2 ms and characteristic sub-pulse widths of 1 ms when examining single pulses from J1818 at observing frequencies between 4 and 8 GHz.

To search for quasi-periodic sub-pulse structure, we first made high-time-resolution timeseries of each observation. Each observation was de-dispersed and written out to a timeseries with 8192 pulse phase bins, equivalent to a 166 $\mu$s time resolution. The sub-pulse structure resolution is lost at 820 MHz due to dispersive smearing of the pulses, and there was an insufficient number of bright pulses at 35 GHz to perform this analysis.

For the remaining observations, we took an on-pulse cutout covering 20\% of the magnetar's rotation (roughly 275 ms in duration) for each single pulse and took its autocorrelation function. For a given pulse timeseries $I(n)$ with $n$ pulse phase bins, the value of the ACF at a time lag $k$ is given by

\begin{equation}
   {\rm ACF}(k) = \sum_{n} I(n+k) * I(n)
\end{equation}

If there is periodic structure within the subpulses, there will be several equally spaced local maxima in the ACF. The first local maximum corresponds to the fundamental quasi-periodicity, and the rest are harmonics located at multiples of the fundamental period (\citealt{psrhandbook,kld+24}). Similarly, the characteristic width of the individual subpulses can be determined by measuring the first local minimum preceding the first local maximum.

\begin{figure}
    \centering
    \includegraphics[scale=0.55]{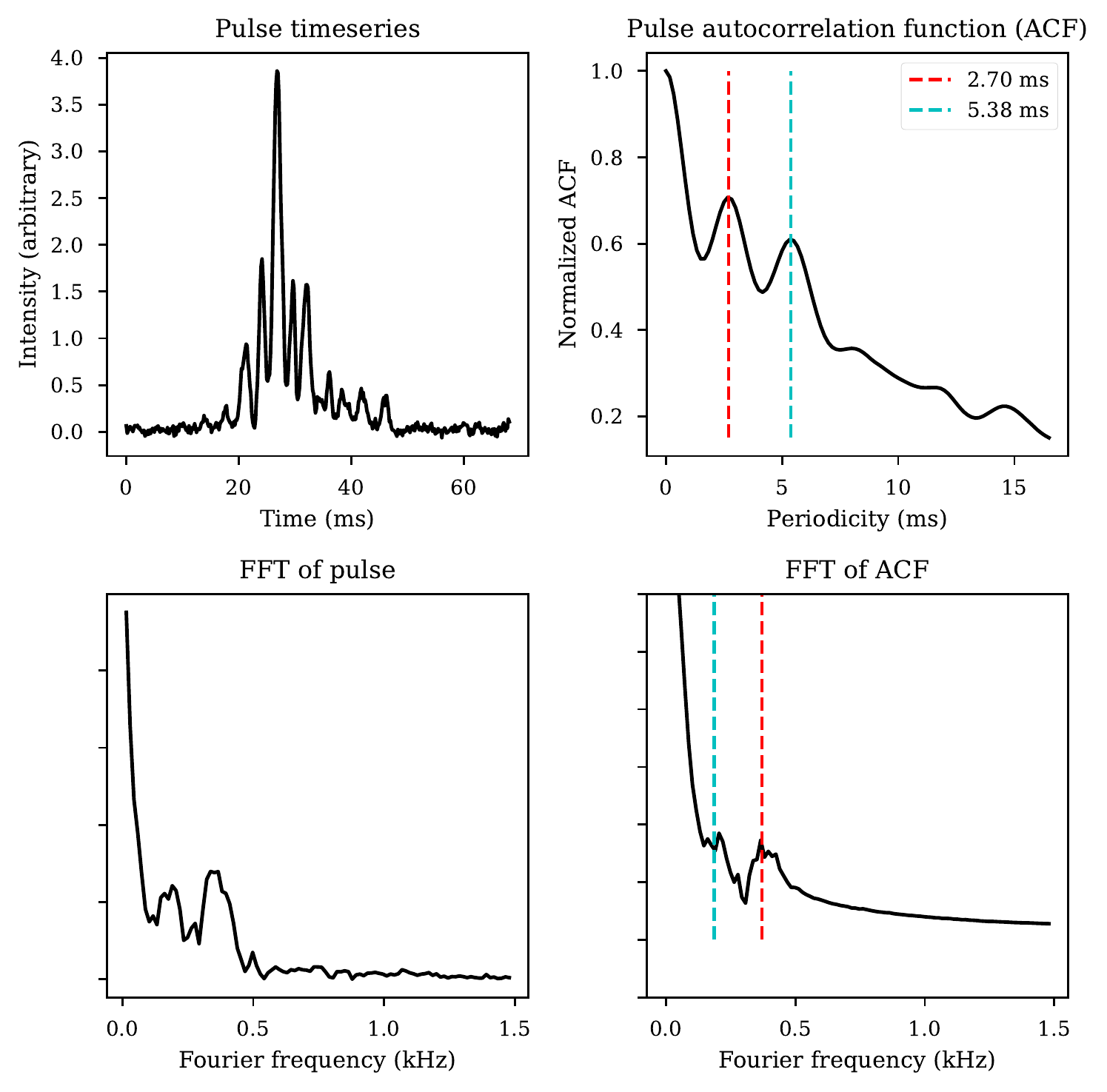}
    \caption{An example autocorrelation function analysis for a pulse at 6 GHz. Top left: pulse timeseries. Top right: pulse autocorrelation function. The identified peaks are denoted with dashed vertical lines. Bottom left: the normalized Fourier transform of the pulse timeseries. Bottom right: the normalized Fourier transform of the ACF. Vertical lines corresponding to the identified quasi-periodicities have been placed at the corresponding Fourier frequencies.}
    \label{fig:acf_example}
\end{figure}

We chose pulses with peak significance $\sigma \geq$ 4 as defined in Section~\ref{sec:pulse_energies} to perform the autocorrelation analysis. For each pulse, we calculated the ACF and the Fourier transform of the ACF, and used the \texttt{scipy.optimize} implementation of cubic spline fitting to search for local maxima in both the ACF and its Fourier transform. An example pulse with a detected quasi-periodicity of 2.7 ms and its ACF analysis are shown in Figure~\ref{fig:acf_example}.

Any identified peaks in the ACF or its Fourier transform are saved. At each frequency, we also examined the 100 single pulses with the highest peak significance by eye, and noted any pulses with clear periodically spaced peaks in the ACF, which the cubic spline fitting did not identify.

For each of these pulses and their associated quasi-periodicities, we also wanted to measure a statistical significance for each quasi-periodicity. We took each pulse with an identified quasi-periodicity and randomly scrambled the order of its phase bins 50000 times. We took the ACF of each scrambled pulse and identified the maximum value outside of the first and last bin (e.g. the quasi-periodicity of the simulated pulse), and summed the values of the ACF at the maximum value and its harmonics. The simulated ACF sums form a null distribution, and the mean and standard deviation of this distribution can be compared to the summed ACF power of the real quasi-periodicity to obtain a significance ($\sigma$) for the detected quasi-periodicity.

\begin{figure}
    \centering
    \includegraphics[scale=0.65]{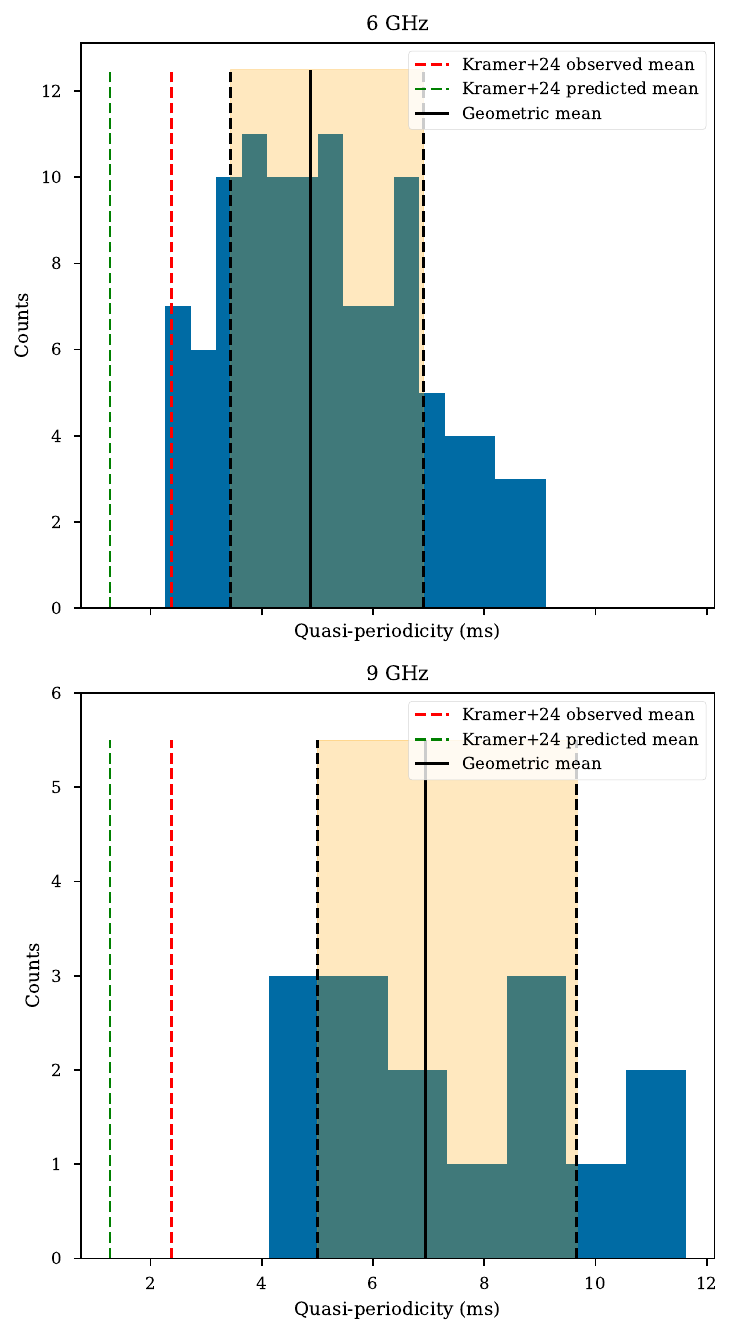}
    \caption{The distribution of detected quasi-periodicities in the single pulses of J1818 at 6 GHz (top) and 9 GHz (bottom). The observed and predicted means of the quasi-periodicities found by \citet{kld+24} are denoted with dashed red and green lines, respectively. The geometric mean and associated error bars for our distribution of detected quasi-periodicities are denoted with black lines and shaded in orange.}
    \label{fig:QP_dists}
\end{figure}

Of the pulses selected for autocorrelation analysis, 108 pulses at 6 GHz and 15 pulses at 9 GHz had detected quasi-periodicities with significances above 4 $\sigma$. The distribution of quasi-periodicities at both frequencies is shown in Figure~\ref{fig:QP_dists}. At 6 GHz, the geometric mean $\mu_g$ of the detected quasi-periodicities is $4.9^{+2.0}_{-1.4}$ ms; at 9 GHz, the mean is $6.9^{+2.7}_{-1.9}$ ms, where the 1-$\sigma$ error bars are calculated using the geometric standard deviation $\sigma_g$; that is, the error bars range from $\mu_g/\sigma_g$ to $\mu_g \times \sigma_g$. The power-law scaling relation derived by \citet{kld+24} predicts a mean quasi-periodicity of $1.27 \pm 0.08$ ms, while the distribution of their detected quasi-periodicities had a geometric mean of $2.37 \pm 0.04$ ms.

The quasi-periodicities detected at 6 GHz by \citet{kld+24} were all shorter than 5 ms, at odds with the longer quasi-periodicities distributed over a larger range which we observe at both frequencies. In both radio pulsars and magnetars, a large spread of detected quasi-periodicity values have been measured (\citealt{mar+15,kld+24}). In particular, a clear bimodal distribution of detected periodicities was reported by \citet{kld+24} for the radio magnetar XTE J1810--197. We find that at both frequencies, the shorter quasi-periodicities have higher statistical significance than the longer quasi-periodicities. If the detection threshold were raised to significances $\geq 8$, the mean periodicities would become $3.8^{+1.4}_{-1.0}$ ms at 6 GHz (from 44 pulses), and $6.2^{+2.1}_{-1.6}$ ms at 9 GHz (from seven pulses). Our results show a clear temporal variation in the periodic substructure of magnetars over time-- regardless of the statistical significance threshold used, the mean quasi-periodicity of our detected pulses is higher than those preferred by \citet{kld+24}.

It is also worth noting that for all of the radio magnetars studied by \citet{kld+24}, the geometric mean of their quasi-periodicities is larger than that predicted by the fitted power-law scaling relationship. While their periodicities still agreed with the scaling relationship within one to two geometric standard deviations, further study would be necessary to determine whether the periodic sub-structure relationship differs for magnetars as opposed to normal pulsars. The periodicity of the sub-structure appears to only depend on the underlying rotational period of the neutron star for a diverse range of neutron star types observed over a wide range of observing frequencies. While our results are an outlier compared to the predictions of the scaling relationship proposed by \citet{kld+24}, a larger sample of pulses over a range of observing epochs would be necessary to draw any conclusions about the nature of quasi-periodic substructure in magnetar single pulses.

\section{Discussion and Conclusions} \label{sec:disc}

\subsection{The Gigahertz-Peaked Spectrum of \swiftj}
The pulsed radio emission from magnetars tends to have a flatter spectrum than that of pulsars, allowing their detection at high observing frequencies. In this section, we discuss our measurement of steep, negative spectral indices above 9 GHz in light of the previously observed spectral properties of J1818 and other radio magnetars, and posit that J1818 exhibits radio emission with a gigahertz-peaked spectrum (GPS). 

\swiftj possessed a steep, negative radio spectral index ($\alpha < -1.8$) for the first few months post-outburst (\citealt{2020ATel13560....1M, lsj+20,ccc+20}), and was not detectable at or above 8 GHz (\citealt{2020ATel13575....1G,2020ATel13587....1L}). Four months post-outburst, a significant spectral flattening was detected, with a spectral index below 9 GHz of $\alpha = +0.3(2)$ \citep{2020ATel13898....1M}. Follow-up observations yielded high-frequency detections of J1818 up to 154 GHz, implying an overall flattening of the spectral index with time \citep{2020ATel14001....1T}. Longer-term followup found that in the months preceding our observations, the spectral index of J1818 between 2 and 8 GHz flattened over time (\citealt{hys+21,bwp+23}). 

It is not uncommon for radio magnetars, including J1818, to emit steeper-spectrum emission at radio frequencies above 10 GHz. For instance, when J1818 was first detected above 8 GHz, fitting a single power law to the flux measurements between observing bands yielded steeper, negative spectral indices (\citealt{2020ATel13966....1P,2020ATel13997....1L,2020ATel14001....1T}). In fact, \citet{hys+21} report an inverted spectral index ($\alpha=+0.5$) between 2 and 8 GHz on the same day that \citet{2020ATel14001....1T} report a negative spectral index ($\alpha=-1.4$) between 86 and 154 GHz, leading all of the above authors to propose that J1818 exhibits a spectral turnover somewhere at high frequencies. 

Based on our flux measurements (Figure~\ref{fig:phase_spectra}), the flux of J1818 increases with frequency up to a peak frequency of $\sim$ 5 GHz, before experiencing a sharp downturn. The instantaneous, in-band spectral indices of all profile components of J1818 are quite steep above 9 GHz, displaying spectral indices steeper than the average value of $\alpha = -1.6$ for radio pulsars \citep{jvk+18}. The clear spectral peak, and steep spectral indices at frequencies above the peak frequency, lead us to classify \swiftj as having a gigahertz-peaked spectrum.

Most radio magnetars have, at least temporarily, shown evidence of radio spectra which peak at gigahertz frequencies (Gigahertz Peaked Spectrum; GPS). These include XTE J1810--197 \citep{eta+21}, SGR 1745--2900 \citep{ppe+15}, 1E 1547.1--5408, and PSR J1622--4950 \citep{ktl+13}. GPS are also observed in a very small subset of rotation-powered radio pulsars, albeit with lower peak frequencies (0.6 to 2 GHz) than the bulk of the radio magnetar population \citep{jvk+18}. The prevailing explanation for these spectral peaks is an interaction between the radio emission and the ionized environments around the neutron star; it is also possible that the younger magnetars are situated in denser or hotter environments, leading to higher peak frequencies (\citealt{ktl+13,kbl+17}).

Diffuse X-ray and radio emission have been detected in the region around J1818. The X-ray emission is indicative of a dust-scattering halo, and a semi-circular region of radio emission was found surrounding the location of J1818 (\citealt{bs20,ibr+23}). Recent VLBA astrometry has determined the proper motion and distance of J1818, and determined that their parallax distance of 9 kpc still allows for the interpretation of the diffuse radio emitting region as a potential supernova remnant association \citep{dld+24}. This radio emitting region could contribute to the observed GPS of J1818; however, even if it is not an associated SNR, other absorbing regions along the line of sight could produce these spectral features as well \citep{kbl+17}. 



\subsection{Pulse Profile Morphology}
Across the frequency range of our observations, the integrated profiles consisted of a narrow, consistent central component (II), flanked by more sporadic components (I and III) comprised of narrow sub-pulses. The outer components tended to be comprised of bright, narrow subpulses and had higher levels of flux modulation in time, similar to the emission properties observed in radio magnetars. Furthermore, the majority of pulses at 9 GHz with the sign of the circular polarization reversed originated from the phase range of component III only. 

Early observations of J1818 found that its integrated profile typically consisted of a single peak with a high degree of linear polarization and a steep spectrum (\citealt{lsj+20,ccc+20}). This profile component was consistently detected across multiple epochs and frequencies, and its properties are similar to those we observe in Component II. Brighter, more sporadic pulses were occasionally detected from a trailing component as early as five days post-outburst (\citealt{2020ATel13560....1M,ccc+20}), suggesting it is the same sporadic trailing component that we observe as Component III. 

We only detect the leading component (I) with high significance at 13 and 22 GHz, the latest two observations in our dataset. 12 pulses with a significance level $\geq 4$ (as defined in Section~\ref{sec:pulse_energies}) were detected from the phase range of Component I in the 9-GHz dataset. A low S/N leading component is visible by eye in the 35 GHz profile, though it is offset in longitude from the position of the leading component at other frequencies. 

Emission from a leading component was also detected by \citet{ljs+21} and \citet{hys+21}. The leading component detected at 2.5 GHz in \citet{ljs+21} seemed to appear sometime between MJDs 59009 and 59047, and over the next two months shifted in pulse longitude until it nearly overlapped the main pulse component at the end of their observations on MJD 59128. Similar drifts in the pulse longitude of individual pulse components were also observed in the radio magnetar XTE J1810--197 \citep{lld+19}. This component may have a sufficiently steep spectrum to not be detectable at lower frequencies, or the longitudinal drift may have reversed direction between observations. 

\citet{hys+21} took simultaneous 2.25- and 8.6-GHz observations of J1818, which ran from MJD 58936 until MJD 59092 (the date of our first observation). In these observations, the middle component is consistent in time and generally narrower at 8.6 GHz than at 2.2 GHz, and the flanking profile components tend to be wider and more sporadic than the middle component. 

\subsubsection{Pulse Widths}
Many rotation-powered pulsars display narrower pulse widths at higher observing frequencies \citep{phs+16}. Pulse width narrowing with increasing frequency was also observed in the case of the magnetar XTE J1810--197 \citep{eta+21}. The evolution of pulse width with observing frequency in canonical pulsars is explained with radius-frequency mapping (RFM), which predicts that pulse widths narrow at higher radio frequencies \citep{cordes78}. The FWHM of the main component of the integrated profile remains mostly consistent across our observing frequencies, yet significantly broadens between 22 and 35 GHz (see Table~\ref{tab:prof_components}). Simultaneous, multi-frequency observations would be necessary in order to determine whether these variations are frequency- or epoch-dependent.

\subsubsection{Mode Switching}

We find that over the course of our observations, the integrated pulse profile of \swiftj was quite stable at all observing frequencies on timescales of minutes. J1818 has previously exhibited mode changing behavior, alternating between quasi-stable emission states on timescales of minutes, but only very occasionally. J1818 seemed to enter a period of enhanced mode-changing activity between 2020 June and 2020 August, when mode switching was observed at frequencies of 1.4 GHz \citep{rsl+22}, and simultaneously at 2.2 and 8.5 GHz (\citealt{hys+21,bwp+23}). \citet{lsj+20} also observe two distinct types of mode switching below 4 GHz, on minutes-long timescales, though the modes have different emission properties than the modes seen in the other studies. 

High-cadence monitoring of J1818 at 1.4 and 6 GHz revealed that J1818 displayed three distinct emission modes during the course of our observing campaign, which provide important context for these results (\citealt{rsl+22,fbr+24}). Our 6-GHz observation on MJD 59092 occurred during Mode I (as labeled by \citet{rsl+22}), when the integrated pulse profile at 1.4 and 6 GHz was  single-peaked. We observe similar behavior in our 6 GHz profile, although we observe a low-intensity trailing component.

The first mode change occurred at around MJD 59100, after which the trailing component appeared, and the flux of the main component became more variable. The rest of our observations took place during Mode II, except for our observation at 13 GHz which took roughly 15 days after the switch to Mode III.

The Mode II observed by \citet{rsl+22} is marked by the narrow main component followed by a broader trailing component; the leading component is barely visible in their 6 GHz profile and not at all visible in their 1.4 GHz profile. In our observations taken during this mode, the leading component is barely visible at 9 GHz, but more prominent at 22 GHz. The mode change to Mode III took place eleven days after our 22-GHz observation according to \citet{fbr+24}, after which point the leading component became much more prominent. Mode changing is usually quasi-instantaneous, and the daily observing cadence of \citet{rsl+22} has localized the epoch of the mode change to within one day, so we cannot relate the appearance of this component with the reported mode change. The profile variability maps shown in \citet{fbr+24} reveal that, the emission from the precursor component was weaker than average before MJD $\sim$59206, and it is not at all visible at 1.4 GHz in the integrated profile taken just a day before our 22 GHz observation. The appearance of the leading component in our integrated profiles before the reported mode change indicates the frequency- dependent evolution of this component, independent of the mode changing phenomenon.




\subsection{Single Pulse Properties}

\subsubsection{Pulse Energies}
The Crab pulsar, among other pulsars, has been observed to emit giant pulses, usually defined as having a flux above 10 times the average pulse flux; pulse amplitude distributions containing giant pulses are best fit with a power-law tail \citep{mml+12}. On the other hand, the pulse energy distributions of many standard pulsars follow a log-normal distribution, with no high-energy tail \citep{bjb+12}. Comparisons in the pulse energy distributions of rotating radio transients (RRATs) and standard pulsars have also been used to clarify their emission mechanisms (\citealt{cbm+17,mmm+18}).

The log-normal distribution provides a better fit for the observations at 820 MHz, 6 GHz, and 9 GHz. At 13 and 22 GHz, the log-normal fits underestimate the number of pulses with fluxes $\gtrsim$ 1.5 times the average flux; however, the small number of pulses in many of the high-flux bins make the use of the chi-squared test less accurate. Given that the normalized fluxes at these frequencies do not surpass five times the average peak flux density, they are likely not created by a giant pulse emission mechanism as seen in other pulsars, e.g. the Crab.

The only pulses resembling giant pulses in our sample are some bright pulses from Component III at 6 GHz. Though these pulses surpass the criterion of a pulse energy $\geq$ 10 times the average energy, at higher frequencies, no pulses pass this threshold, and the power-law fit only outperforms a log-normal fit at 6 and 9 GHz. Interestingly, the pulse energy distributions from Component II are better fit by power law distributions than log-normal distributions, except for at 9 GHz where the model slightly overfits the data (e.g., $\chi^2_r < 1$). In all other regards, the central component has tended to display more pulsar-like characteristics, so it would be a reasonable expectation that the log-normal fits would be preferred for the central component's energy distributions. 

The pulse energy distributions of the Galactic center magnetar SGR J1745--2900 have been well-fit by log-normal distributions before (\citealt{lak+15,ywm+18}). A similar lack of a high-energy tail from the single pulses of J1818 was observed by \citet{lsj+20}, and \citet{ccc+20} detected only one pulse with an energy $>$ 10 times the average, a few weeks after outburst. A study of pulse energy distributions from the magnetar XTE J1810--197 between 1--8 GHz found that the energies are not well-described by a single statistical distribution, and varied on timescales of days \citep{ssw+09}. 

\subsubsection{Frequency Structure}
Other radio magnetar studies have found frequency structure in magnetar bursts, similar to fast radio bursts (\citealt{pmp+18,mjs+19}). In these cases, individual sub-pulses within a magnetar rotation would have flux densities which varied by factors of 2--10 across the observing band, as well as the radio emission disappearing over $\sim$ 100-MHz wide intervals in the observing band. As exemplified in Figure~\ref{fig:C_sps}, the single pulses from our observations of J1818 displayed broadband emission, and we did not observe narrowband frequency structure in our single pulses, nor did we observe the downward drifting frequency behavior in subsequent subpulses often seen in repeating FRBs. \citet{mjs+19} showed that in the case of XTE J1810--197, high degrees of frequency modulation were more common closer to the beginning of the outburst, and that the frequency structure of the magnetar single pulses tended to be more uniform at later observing dates. 

\section{Conclusions}
\label{sec:conclusion}
We carried out observations of the radio magnetar \swiftj with the GBT over a range of frequencies from 0.8 to 35 GHz beginning five months after its March 2020 outburst. At observing frequencies above 6 GHz, the spectral index of the radio emission overturns and significantly steepens, indicating that J1818 at least temporarily exhibited a gigahertz-peaked spectrum (GPS) commonly seen in radio magnetars. A potential supernova remnant has previously been identified in the region around J1818 in the form of a semi-circular radio-emitting region and extended dust-scattering X-ray halo, which could cause the observed GPS.

The integrated pulse profile of J1818 was also remarkably stable across frequencies, consisting of up to three pulse components. The central component (Component II) tends to be narrower and displays less flux variability over time when compared to the other two components. The trailing component (Component III) in particular is composed largely of narrow, bright subpulses which fall within the larger pulse envelope, reminiscient of the highly variable radio bursts from magnetars. At most frequencies, the outer profile components have flatter spectral indices and higher flux modulation indices than the central component.  

We did not observe FRB-like behavior from J1818 in these observations, namely, narrowband pulses with downward-drifting frequency structure, nor pulses with sufficient energies to be detected as FRBs had J1818 been more distant. Given the stable pulse profile across observing epochs and frequencies, and the lack of mode changing on short timescales, the magnetosphere of J1818 may have settled to a more stable configuration in the months following its initial outburst.

\begin{acknowledgments}
We thank the anonymous reviewer for their comments which greatly improved the manuscript. E.F.L. and M.A.M. are supported by NSF award AST-2009425. M.A.M. is also supported by NSF Physics Frontiers Center award PHYS-2020265. The Green Bank Observatory is a facility of the National Science Foundation operated under cooperative agreement by Associated Universities, Inc. 
\end{acknowledgments}

\facilities{GBT (VEGAS)}

\software{
\texttt{Astropy} \citep[\url{https://www.astropy.org/};][]{astropy2022},
\texttt{Matplotlib} \citep[\url{https://matplotlib.org/};][]{matplotlib},
\texttt{Numpy} \citep[\url{https://numpy.org/};][]{numpy},
\texttt{PRESTO} \citep[\url{https://www.cv.nrao.edu/~sransom/presto/};][]{presto}, 
\texttt{PSRCHIVE} \citep[\url{https://psrchive.sourceforge.net/};][]{hsm+04, psrchive},
\texttt{pypulse} \citep[\url{https://github.com/mtlam/PyPulse};][]{pypulse},
\texttt{pyGDSM} \citep[\url{https://github.com/telegraphic/pygdsm};][]{2016ascl.soft03013P},
\texttt{Scipy} \citep[\url{https://scipy.org/};][]{scipy},
\texttt{seaborn} \citep[\url{https://github.com/mwaskom/seaborn};][]{seaborn},
\texttt{TEMPO2} \citep[\url{https://www.atnf.csiro.au/research/pulsar/tempo2/};][]{tempo2}
}

\bibliography{biblio}{}
\bibliographystyle{aasjournal}

\end{document}

%% file: observations.tex
\begin{deluxetable}{ccccccc}
\label{tab:obs}
\tablecaption{Observations}
\tablehead{\colhead{Start Date} & \colhead{MJD} & \colhead{$F_c$} & \colhead{$\Delta\nu$} & \colhead{$N_{chan}$} & \colhead{$t_{samp}$} & \colhead{Length} \\
 & & \colhead{(GHz)} & \colhead{(GHz)} & & \colhead{($\mu$s)} & \colhead{(hr)}}
\startdata
2020 Aug 31 & 59092 & 6.00 & 4.50 & 3072 & 87.38 & 3.4 \\
2020 Sep 27 & 59119 & 35.12 & 3.75 & 5118 & 87.38 & 1.2 \\
2020 Oct 29 & 59151 & 0.82 & 0.20 & 128 & 10.24 & 1.6 \\
2020 Oct 30 & 59152 & 9.11 & 2.25 & 1536 & 87.38 & 1.1 \\
2020 Nov 20 & 59173 & 21.94 & 7.87 & 5376 & 87.38 & 1.4 \\
2020 Dec 07 & 59190 & 13.69 & 3.37 & 2304 & 87.38 & 1.4 \\
\enddata
\tablecomments{Parameters for each observation of \swiftj as described in the text. $F_c$ denotes the central frequency of the observation, $\Delta\nu$ the bandwidth, $N_{chan}$ the total number of frequency channels, and $t_{samp}$ the native sampling time.}
\end{deluxetable}

%% file: prof_components.tex
\begin{deluxetable*}{c|ccc|ccc|ccc|ccc}
\tablecaption{Pulse component properties}
\tablehead{
\colhead{Frequency} & \multicolumn{3}{c}{Pulse Width (FWHM)} & \multicolumn{3}{c}{$\alpha$} & \multicolumn{3}{c}{\textit{L/I}}  & \multicolumn{3}{c}{\textit{C/I}} \\
\colhead{(GHz)}     & \multicolumn{3}{c}{(ms)}  & \multicolumn{3}{c}{}         & \multicolumn{3}{c}{(\%)} & \multicolumn{3}{c}{(\%)} \\
\hline
\multicolumn{1}{c|}{Component} & I & II & III & I & II & III & I & II & III & I & II & III 
}
\startdata
0.82 & \multicolumn{3}{c|}{\nodata} & \multicolumn{3}{c|}{\nodata} & \multicolumn{3}{c|}{54.5(7)} & \multicolumn{3}{c}{9(3)}  \\
6.00 & (a) & 16.0(1) & 33.4(4) & (a) & (b) & (b) & & (c) & & & (c) & \\
9.11 & (a) & 15.8(2) & 74.9(6) & (a) & $-$3.3(5) & $-$2.6(4) & (a) & 60(10) & 20(11) & (a) & 29(9) & \nodata \\
13.69 & 37.6(5) & 15.9(1) & 65.0(3) & $-$2(2) & $-$2.3(6) & $-$2.1(4) & & (c) & & & (c) & \\
21.94 & 23.6(4) & 17.4(6) & 63(1) & $-$3(2) & $-$3.4(8) & $-$3.4(6) & 70(15) & 60(5) & 65(3) & \nodata & 25(17) & 20(17) \\
35.12 & (a) & 34(2) & 59(1) & \multicolumn{3}{c|}{\nodata} & (a) & \nodata & 42(16) & (a) & \nodata & \nodata \\
\enddata
\tablecomments{Fitted full-width half-maximum (FWHM) pulse widths, spectral indices, linear polarization fractions, and circular polarization fractions for the three profile components depicted in Figure~\ref{fig:multifreq_profs}. The 1$-\sigma$ error bars are denoted by parentheses on the last significant digit. We only present values for profile components which are detected significantly enough to reliably measure these properties. The 820-MHz profile is not divided into distinct components: the linear and polarization fractions are calculated over the entire range of the pulse profile. }
\label{tab:prof_components}
\tablenotetext{a}{Profile component not significantly detected.}
\tablenotetext{b}{Not well-fit by a simple power law; see Section~\ref{sec:spectrum}.}
\tablenotetext{c}{Polarization calibration not available for this observation.}
\end{deluxetable*}

%% file: spidx.tex
\begin{deluxetable}{ccccc}
\label{tab:spidx}
\tablecaption{Spectral indices}
\tablehead{\colhead{Date} & \colhead{$F_c$} & \colhead{$S_{mean}$} & \colhead{Freq range} & \colhead{$\alpha$} \\
\colhead{(MJD)} & \colhead{(GHz)} & \colhead{(mJy)} & \colhead{(GHz)} & \colhead{(in band)}}
\startdata
59151 & 0.82 & 0.359(4) & (0.76, 0.92) & +0.2(1) \\
59092 & 6.0 & 2.56(1) & \nodata & \nodata \\
59152 & 9.1 & 1.35(4) & (8.3, 10.2) & $-$2.9(3) \\
59190 & 13.7 & 0.58(2) & (12.0, 15.3) & $-$2.2(2) \\
59173 & 21.9 & \textbf{0.9(1)} & (18.0, 25.9) & $-$3.2(4) \\
59119 & 35.1 & 0.49(6) & \nodata & \nodata \\
\enddata
\tablecomments{The central frequency ($F_c$) of each observation along with $S_{mean}$, the phase-averaged mean flux at that frequency. The 1-$\sigma$ error bars are denoted with parentheses on the last significant digit. The spectral indices from the fits in Figure~ \ref{fig:spectra_grid} are also listed along the frequency ranges over which they are measured.}
\end{deluxetable}

%% file: spstats_with_flux.tex
\begin{deluxetable}{cccccc}
\label{tab:spstats}
\tablecaption{Single pulse statistics}
\tablehead{\colhead{Date} & \colhead{$F_c$} & \colhead{$N_{rot}$} & \colhead{$N_{pulse}$} & \colhead{$\beta$} & \colhead{$S_{peak,mean}$}\\
\colhead{(MJD)} & \colhead{(GHz)} & & & \colhead{($\mathrm{hr}^{-1}$)} & \colhead{(mJy)}}
\startdata
59151 & 0.82 & 4269 & 2314 & 1450 & 4.7 \\
59092 & 6.0 & 9086 & 8527 & 2500 & 93 \\
59152 & 9.1 & 3022 & 1261 & 1150 & 101 \\
59190 & 13.7 & 3653 & 675 & 480 & 72 \\
59173 & 21.9 & 3745 & 1282 & 920 & 17 \\
59119 & 35.1 & 3251 & 10 & $\sim$8 & \nodata \\
\enddata
\tablecomments{The central frequency ($F_c$) of each observation along with the number of magnetar rotations covered by our observations ($N_{rot}$), the number of pulses detected with significance $\sigma \geq 4$ ($N_{pulse}$), the average detected burst rate ($\beta$), and the mean peak flux of all detected ($\sigma \geq 4$) pulses on that epoch. There are too few bright pulses at 35 GHz for a meaningful average.} 
\end{deluxetable}

%% file: main.bbl
\begin{thebibliography}{}
\expandafter\ifx\csname natexlab\endcsname\relax\def\natexlab#1{#1}\fi
\providecommand{\url}[1]{\href{#1}{#1}}
\providecommand{\dodoi}[1]{doi:~\href{http://doi.org/#1}{\nolinkurl{#1}}}
\providecommand{\doeprint}[1]{\href{http://ascl.net/#1}{\nolinkurl{http://ascl.net/#1}}}
\providecommand{\doarXiv}[1]{\href{https://arxiv.org/abs/#1}{\nolinkurl{https://arxiv.org/abs/#1}}}

\bibitem[{{Archibald} {et~al.}(2016){Archibald}, {Kaspi}, {Tendulkar}, \& {Scholz}}]{akt+16}
{Archibald}, R.~F., {Kaspi}, V.~M., {Tendulkar}, S.~P., \& {Scholz}, P. 2016, \apjl, 829, L21, \dodoi{10.3847/2041-8205/829/1/L21}

\bibitem[{{Astropy Collaboration} {et~al.}(2022){Astropy Collaboration}, {Price-Whelan}, {Lim}, {Earl}, {Starkman}, {Bradley}, {Shupe}, {Patil}, {Corrales}, {Brasseur}, {N{\"o}the}, {Donath}, {Tollerud}, {Morris}, {Ginsburg}, {Vaher}, {Weaver}, {Tocknell}, {Jamieson}, {van Kerkwijk}, {Robitaille}, {Merry}, {Bachetti}, {G{\"u}nther}, {Aldcroft}, {Alvarado-Montes}, {Archibald}, {B{\'o}di}, {Bapat}, {Barentsen}, {Baz{\'a}n}, {Biswas}, {Boquien}, {Burke}, {Cara}, {Cara}, {Conroy}, {Conseil}, {Craig}, {Cross}, {Cruz}, {D'Eugenio}, {Dencheva}, {Devillepoix}, {Dietrich}, {Eigenbrot}, {Erben}, {Ferreira}, {Foreman-Mackey}, {Fox}, {Freij}, {Garg}, {Geda}, {Glattly}, {Gondhalekar}, {Gordon}, {Grant}, {Greenfield}, {Groener}, {Guest}, {Gurovich}, {Handberg}, {Hart}, {Hatfield-Dodds}, {Homeier}, {Hosseinzadeh}, {Jenness}, {Jones}, {Joseph}, {Kalmbach}, {Karamehmetoglu}, {Ka{\l}uszy{\'n}ski}, {Kelley}, {Kern}, {Kerzendorf}, {Koch}, {Kulumani}, {Lee}, {Ly}, {Ma}, {MacBride}, {Maljaars}, {Muna}, {Murphy}, {Norman},
  {O'Steen}, {Oman}, {Pacifici}, {Pascual}, {Pascual-Granado}, {Patil}, {Perren}, {Pickering}, {Rastogi}, {Roulston}, {Ryan}, {Rykoff}, {Sabater}, {Sakurikar}, {Salgado}, {Sanghi}, {Saunders}, {Savchenko}, {Schwardt}, {Seifert-Eckert}, {Shih}, {Jain}, {Shukla}, {Sick}, {Simpson}, {Singanamalla}, {Singer}, {Singhal}, {Sinha}, {Sip{\H{o}}cz}, {Spitler}, {Stansby}, {Streicher}, {{\v{S}}umak}, {Swinbank}, {Taranu}, {Tewary}, {Tremblay}, {de Val-Borro}, {Van Kooten}, {Vasovi{\'c}}, {Verma}, {de Miranda Cardoso}, {Williams}, {Wilson}, {Winkel}, {Wood-Vasey}, {Xue}, {Yoachim}, {Zhang}, {Zonca}, \& {Astropy Project Contributors}}]{astropy2022}
{Astropy Collaboration}, {Price-Whelan}, A.~M., {Lim}, P.~L., {et~al.} 2022, \apj, 935, 167, \dodoi{10.3847/1538-4357/ac7c74}

\bibitem[{{Bansal} {et~al.}(2023){Bansal}, {Wharton}, {Pearlman}, {Majid}, {Prince}, {Younes}, {Hu}, {Enoto}, {Kocz}, \& {Horiuchi}}]{bwp+23}
{Bansal}, K., {Wharton}, R.~S., {Pearlman}, A.~B., {et~al.} 2023, \mnras, 523, 2401, \dodoi{10.1093/mnras/stad1520}

\bibitem[{{Blumer} \& {Safi-Harb}(2020)}]{bs20}
{Blumer}, H., \& {Safi-Harb}, S. 2020, \apjl, 904, L19, \dodoi{10.3847/2041-8213/abc6a2}

\bibitem[{{Bochenek} {et~al.}(2020){Bochenek}, {Ravi}, {Belov}, {Hallinan}, {Kocz}, {Kulkarni}, \& {McKenna}}]{brb+20}
{Bochenek}, C.~D., {Ravi}, V., {Belov}, K.~V., {et~al.} 2020, \nat, 587, 59, \dodoi{10.1038/s41586-020-2872-x}

\bibitem[{{Burke-Spolaor} {et~al.}(2012){Burke-Spolaor}, {Johnston}, {Bailes}, {Bates}, {Bhat}, {Burgay}, {Champion}, {D'Amico}, {Keith}, {Kramer}, {Levin}, {Milia}, {Possenti}, {Stappers}, \& {van Straten}}]{bjb+12}
{Burke-Spolaor}, S., {Johnston}, S., {Bailes}, M., {et~al.} 2012, \mnras, 423, 1351, \dodoi{10.1111/j.1365-2966.2012.20998.x}

\bibitem[{{Camilo} {et~al.}(2008){Camilo}, {Reynolds}, {Johnston}, {Halpern}, \& {Ransom}}]{crj+08}
{Camilo}, F., {Reynolds}, J., {Johnston}, S., {Halpern}, J.~P., \& {Ransom}, S.~M. 2008, \apj, 679, 681, \dodoi{10.1086/587054}

\bibitem[{{Camilo} {et~al.}(2016){Camilo}, {Ransom}, {Halpern}, {Alford}, {Cognard}, {Reynolds}, {Johnston}, {Sarkissian}, \& {van Straten}}]{crh+16}
{Camilo}, F., {Ransom}, S.~M., {Halpern}, J.~P., {et~al.} 2016, \apj, 820, 110, \dodoi{10.3847/0004-637X/820/2/110}

\bibitem[{{Champion} {et~al.}(2020){Champion}, {Cognard}, {Cruces}, {Desvignes}, {Jankowski}, {Karuppusamy}, {Keith}, {Kouveliotou}, {Kramer}, {Liu}, {Lyne}, {Mickaliger}, {O'Connor}, {Parthasarathy}, {Porayko}, {Rajwade}, {Stappers}, {Torne}, {van der Horst}, \& {Weltevrede}}]{ccc+20}
{Champion}, D., {Cognard}, I., {Cruces}, M., {et~al.} 2020, \mnras, 498, 6044, \dodoi{10.1093/mnras/staa2764}

\bibitem[{{CHIME/FRB Collaboration} {et~al.}(2020){CHIME/FRB Collaboration}, {Andersen}, {Bandura}, {Bhardwaj}, {Bij}, {Boyce}, {Boyle}, {Brar}, {Cassanelli}, {Chawla}, {Chen}, {Cliche}, {Cook}, {Cubranic}, {Curtin}, {Denman}, {Dobbs}, {Dong}, {Fandino}, {Fonseca}, {Gaensler}, {Giri}, {Good}, {Halpern}, {Hill}, {Hinshaw}, {H{\"o}fer}, {Josephy}, {Kania}, {Kaspi}, {Landecker}, {Leung}, {Li}, {Lin}, {Masui}, {McKinven}, {Mena-Parra}, {Merryfield}, {Meyers}, {Michilli}, {Milutinovic}, {Mirhosseini}, {M{\"u}nchmeyer}, {Naidu}, {Newburgh}, {Ng}, {Patel}, {Pen}, {Pinsonneault-Marotte}, {Pleunis}, {Quine}, {Rafiei-Ravandi}, {Rahman}, {Ransom}, {Renard}, {Sanghavi}, {Scholz}, {Shaw}, {Shin}, {Siegel}, {Singh}, {Smegal}, {Smith}, {Stairs}, {Tan}, {Tendulkar}, {Tretyakov}, {Vanderlinde}, {Wang}, {Wulf}, \& {Zwaniga}}]{chime_frb_mag}
{CHIME/FRB Collaboration}, {Andersen}, B.~C., {Bandura}, K.~M., {et~al.} 2020, \nat, 587, 54, \dodoi{10.1038/s41586-020-2863-y}

\bibitem[{{Cordes}(1978)}]{cordes78}
{Cordes}, J.~M. 1978, \apj, 222, 1006, \dodoi{10.1086/156218}

\bibitem[{{Cordes} \& {Lazio}(2002)}]{ne2001}
{Cordes}, J.~M., \& {Lazio}, T.~J.~W. 2002, arXiv e-prints, astro.
\newblock \doarXiv{astro-ph/0207156}

\bibitem[{{Cui} {et~al.}(2017){Cui}, {Boyles}, {McLaughlin}, \& {Palliyaguru}}]{cbm+17}
{Cui}, B.~Y., {Boyles}, J., {McLaughlin}, M.~A., \& {Palliyaguru}, N. 2017, \apj, 840, 5, \dodoi{10.3847/1538-4357/aa6aa9}

\bibitem[{{Dai} {et~al.}(2018){Dai}, {Johnston}, {Weltevrede}, {Kerr}, {Burgay}, {Esposito}, {Israel}, {Possenti}, {Rea}, \& {Sarkissian}}]{djw+18}
{Dai}, S., {Johnston}, S., {Weltevrede}, P., {et~al.} 2018, \mnras, 480, 3584, \dodoi{10.1093/mnras/sty2063}

\bibitem[{{Dai} {et~al.}(2019){Dai}, {Lower}, {Bailes}, {Camilo}, {Halpern}, {Johnston}, {Kerr}, {Reynolds}, {Sarkissian}, \& {Scholz}}]{dlb+19}
{Dai}, S., {Lower}, M.~E., {Bailes}, M., {et~al.} 2019, \apjl, 874, L14, \dodoi{10.3847/2041-8213/ab0e7a}

\bibitem[{{Desvignes} {et~al.}(2024){Desvignes}, {Weltevrede}, {Gao}, {Jones}, {Kramer}, {Caleb}, {Karuppusamy}, {Levin}, {Liu}, {Lyne}, {Shao}, {Stappers}, \& {P{\'e}tri}}]{dwg+24}
{Desvignes}, G., {Weltevrede}, P., {Gao}, Y., {et~al.} 2024, Nature Astronomy, 8, 617, \dodoi{10.1038/s41550-024-02226-7}

\bibitem[{{Ding} {et~al.}(2024){Ding}, {Lower}, {Deller}, {Shannon}, {Camilo}, \& {Sarkissian}}]{dld+24}
{Ding}, H., {Lower}, M.~E., {Deller}, A.~T., {et~al.} 2024, \apjl, 971, L13, \dodoi{10.3847/2041-8213/ad5550}

\bibitem[{{Duncan} \& {Thompson}(1992)}]{dt92}
{Duncan}, R.~C., \& {Thompson}, C. 1992, \apjl, 392, L9, \dodoi{10.1086/186413}

\bibitem[{{Eie} {et~al.}(2021){Eie}, {Terasawa}, {Akahori}, {Oyama}, {Hirota}, {Yonekura}, {Enoto}, {Sekido}, {Takefuji}, {Misawa}, {Tsuchiya}, {Kisaka}, {Aoki}, \& {Honma}}]{eta+21}
{Eie}, S., {Terasawa}, T., {Akahori}, T., {et~al.} 2021, \pasj, 73, 1563, \dodoi{10.1093/pasj/psab098}

\bibitem[{{Enoto} {et~al.}(2020){Enoto}, {Sakamoto}, {Younes}, {Hu}, {Ho}, {Gendreau}, {Arzoumanian}, {Guver}, {Guillot}, {Altamirano}, {Ray}, {Ng}, {Chakrabarty}, {Jaisawal}, \& {Bogdanov}}]{2020ATel13551....1E}
{Enoto}, T., {Sakamoto}, T., {Younes}, G., {et~al.} 2020, The Astronomer's Telegram, 13551, 1

\bibitem[{{Evans} {et~al.}(2020){Evans}, {Gropp}, {Kennea}, {Klingler}, {Laha}, {Lien}, {Page}, {Sakamoto}, {Tohuvavohu}, \& {Neil Gehrels Swift Observatory Team}}]{2020GCN.27373....1E}
{Evans}, P.~A., {Gropp}, J.~D., {Kennea}, J.~A., {et~al.} 2020, GRB Coordinates Network, 27373, 1

\bibitem[{{Fisher} {et~al.}(2024){Fisher}, {Butterworth}, {Rajwade}, {Stappers}, {Desvignes}, {Karuppusamy}, {Kramer}, {Liu}, {Lyne}, {Mickaliger}, {Shaw}, \& {Weltevrede}}]{fbr+24}
{Fisher}, R., {Butterworth}, E.~M., {Rajwade}, K.~M., {et~al.} 2024, \mnras, 528, 3833, \dodoi{10.1093/mnras/stae271}

\bibitem[{{Gajjar} {et~al.}(2020){Gajjar}, {Perez}, {Siemion}, {MacMahon}, {Lebofsky}, {Croft}, \& {Price}}]{2020ATel13575....1G}
{Gajjar}, V., {Perez}, K., {Siemion}, A., {et~al.} 2020, The Astronomer's Telegram, 13575, 1

\bibitem[{Harris {et~al.}(2020)Harris, Millman, van~der Walt, Gommers, Virtanen, Cournapeau, Wieser, Taylor, Berg, Smith, Kern, Picus, Hoyer, van Kerkwijk, Brett, Haldane, del R{\'{i}}o, Wiebe, Peterson, G{\'{e}}rard-Marchant, Sheppard, Reddy, Weckesser, Abbasi, Gohlke, \& Oliphant}]{numpy}
Harris, C.~R., Millman, K.~J., van~der Walt, S.~J., {et~al.} 2020, Nature, 585, 357, \dodoi{10.1038/s41586-020-2649-2}

\bibitem[{{Hobbs} {et~al.}(2006){Hobbs}, {Edwards}, \& {Manchester}}]{tempo2}
{Hobbs}, G.~B., {Edwards}, R.~T., \& {Manchester}, R.~N. 2006, \mnras, 369, 655, \dodoi{10.1111/j.1365-2966.2006.10302.x}

\bibitem[{{Hotan} {et~al.}(2004){Hotan}, {van Straten}, \& {Manchester}}]{hsm+04}
{Hotan}, A.~W., {van Straten}, W., \& {Manchester}, R.~N. 2004, \pasa, 21, 302, \dodoi{10.1071/AS04022}

\bibitem[{{Hu} {et~al.}(2020){Hu}, {Begi{\c{c}}arslan}, {G{\"u}ver}, {Enoto}, {Younes}, {Sakamoto}, {Ray}, {Strohmayer}, {Guillot}, {Arzoumanian}, {Palmer}, {Gendreau}, {Malacaria}, {Wadiasingh}, {Jaisawal}, \& {Majid}}]{hbg+20}
{Hu}, C.-P., {Begi{\c{c}}arslan}, B., {G{\"u}ver}, T., {et~al.} 2020, \apj, 902, 1, \dodoi{10.3847/1538-4357/abb3c9}

\bibitem[{{Huang} {et~al.}(2021){Huang}, {Yan}, {Shen}, {Tong}, {Lin}, {Yuan}, {Liu}, {Zhao}, {Ge}, \& {Wang}}]{hys+21}
{Huang}, Z.-P., {Yan}, Z., {Shen}, Z.-Q., {et~al.} 2021, \mnras, 505, 1311, \dodoi{10.1093/mnras/stab1362}

\bibitem[{Hunter(2007)}]{matplotlib}
Hunter, J.~D. 2007, Computing in Science \& Engineering, 9, 90, \dodoi{10.1109/MCSE.2007.55}

\bibitem[{{Ibrahim} {et~al.}(2023){Ibrahim}, {Borghese}, {Rea}, {Coti Zelati}, {Parent}, {Russell}, {Ascenzi}, {Sathyaprakash}, {G{\"o}tz}, {Mereghetti}, {Topinka}, {Rigoselli}, {Savchenko}, {Campana}, {Israel}, {Tiengo}, {Perna}, {Turolla}, {Zane}, {Esposito}, {Rodr{\'\i}guez Castillo}, {Graber}, {Possenti}, {Dehman}, {Ronchi}, \& {Loru}}]{ibr+23}
{Ibrahim}, A.~Y., {Borghese}, A., {Rea}, N., {et~al.} 2023, \apj, 943, 20, \dodoi{10.3847/1538-4357/aca528}

\bibitem[{{Jankowski} {et~al.}(2018){Jankowski}, {van Straten}, {Keane}, {Bailes}, {Barr}, {Johnston}, \& {Kerr}}]{jvk+18}
{Jankowski}, F., {van Straten}, W., {Keane}, E.~F., {et~al.} 2018, \mnras, 473, 4436, \dodoi{10.1093/mnras/stx2476}

\bibitem[{{Karuppusamy} {et~al.}(2020){Karuppusamy}, {Desvignes}, {Kramer}, {Porayko}, {Champion}, {Torne}, {Stappers}, {van der Horst}, {Kouveliotou}, \& {O'Connor}}]{2020ATel13553....1K}
{Karuppusamy}, R., {Desvignes}, G., {Kramer}, M., {et~al.} 2020, The Astronomer's Telegram, 13553, 1

\bibitem[{{Kaspi} \& {Beloborodov}(2017)}]{kb17}
{Kaspi}, V.~M., \& {Beloborodov}, A.~M. 2017, \araa, 55, 261, \dodoi{10.1146/annurev-astro-081915-023329}

\bibitem[{{Kijak} {et~al.}(2021){Kijak}, {Basu}, {Lewandowski}, \& {Ro{\.z}ko}}]{kbl+21}
{Kijak}, J., {Basu}, R., {Lewandowski}, W., \& {Ro{\.z}ko}, K. 2021, \apj, 923, 211, \dodoi{10.3847/1538-4357/ac3082}

\bibitem[{{Kijak} {et~al.}(2017){Kijak}, {Basu}, {Lewandowski}, {Ro{\.z}ko}, \& {Dembska}}]{kbl+17}
{Kijak}, J., {Basu}, R., {Lewandowski}, W., {Ro{\.z}ko}, K., \& {Dembska}, M. 2017, \apj, 840, 108, \dodoi{10.3847/1538-4357/aa6ff2}

\bibitem[{{Kijak} {et~al.}(2013){Kijak}, {Tarczewski}, {Lewandowski}, \& {Melikidze}}]{ktl+13}
{Kijak}, J., {Tarczewski}, L., {Lewandowski}, W., \& {Melikidze}, G. 2013, \apj, 772, 29, \dodoi{10.1088/0004-637X/772/1/29}

\bibitem[{{Kramer} \& {Johnston}(2008)}]{kj08}
{Kramer}, M., \& {Johnston}, S. 2008, \mnras, 390, 87, \dodoi{10.1111/j.1365-2966.2008.13780.x}

\bibitem[{{Kramer} {et~al.}(2024){Kramer}, {Liu}, {Desvignes}, {Karuppusamy}, \& {Stappers}}]{kld+24}
{Kramer}, M., {Liu}, K., {Desvignes}, G., {Karuppusamy}, R., \& {Stappers}, B.~W. 2024, Nature Astronomy, 8, 230, \dodoi{10.1038/s41550-023-02125-3}

\bibitem[{{Kramer} {et~al.}(2007){Kramer}, {Stappers}, {Jessner}, {Lyne}, \& {Jordan}}]{ksj+07}
{Kramer}, M., {Stappers}, B.~W., {Jessner}, A., {Lyne}, A.~G., \& {Jordan}, C.~A. 2007, \mnras, 377, 107, \dodoi{10.1111/j.1365-2966.2007.11622.x}

\bibitem[{{Lam}(2017)}]{pypulse}
{Lam}, M.~T. 2017, {PyPulse: PSRFITS handler}, Astrophysics Source Code Library, record ascl:1706.011

\bibitem[{{Levin} {et~al.}(2012){Levin}, {Bailes}, {Bates}, {Bhat}, {Burgay}, {Burke-Spolaor}, {D'Amico}, {Johnston}, {Keith}, {Kramer}, {Milia}, {Possenti}, {Stappers}, \& {van Straten}}]{lbb+12}
{Levin}, L., {Bailes}, M., {Bates}, S.~D., {et~al.} 2012, \mnras, 422, 2489, \dodoi{10.1111/j.1365-2966.2012.20807.x}

\bibitem[{{Levin} {et~al.}(2019){Levin}, {Lyne}, {Desvignes}, {Eatough}, {Karuppusamy}, {Kramer}, {Mickaliger}, {Stappers}, \& {Weltevrede}}]{lld+19}
{Levin}, L., {Lyne}, A.~G., {Desvignes}, G., {et~al.} 2019, \mnras, 488, 5251, \dodoi{10.1093/mnras/stz2074}

\bibitem[{{Liu} {et~al.}(2020){Liu}, {Karuppusamy}, {Cognard}, {Desvignes}, {Kramer}, {Lyne}, {Rajwade}, {Stappers}, \& {Torne}}]{2020ATel13997....1L}
{Liu}, K., {Karuppusamy}, R., {Cognard}, I., {et~al.} 2020, The Astronomer's Telegram, 13997, 1

\bibitem[{{Lorimer} \& {Kramer}(2012)}]{psrhandbook}
{Lorimer}, D.~R., \& {Kramer}, M. 2012, {Handbook of Pulsar Astronomy}

\bibitem[{{Lower} {et~al.}(2021){Lower}, {Johnston}, {Shannon}, {Bailes}, \& {Camilo}}]{ljs+21}
{Lower}, M.~E., {Johnston}, S., {Shannon}, R.~M., {Bailes}, M., \& {Camilo}, F. 2021, \mnras, 502, 127, \dodoi{10.1093/mnras/staa3789}

\bibitem[{{Lower} \& {Shannon}(2020)}]{2020ATel13587....1L}
{Lower}, M.~E., \& {Shannon}, R.~M. 2020, The Astronomer's Telegram, 13587, 1

\bibitem[{{Lower} {et~al.}(2020){Lower}, {Shannon}, {Johnston}, \& {Bailes}}]{lsj+20}
{Lower}, M.~E., {Shannon}, R.~M., {Johnston}, S., \& {Bailes}, M. 2020, \apjl, 896, L37, \dodoi{10.3847/2041-8213/ab9898}

\bibitem[{{Lower} {et~al.}(2023){Lower}, {Younes}, {Scholz}, {Camilo}, {Dunn}, {Johnston}, {Enoto}, {Sarkissian}, {Reynolds}, {Palmer}, {Arzoumanian}, {Baring}, {Gendreau}, {G{\"o}{\u{g}}{\"u}{\c{s}}}, {Guillot}, {van der Horst}, {Hu}, {Kouveliotou}, {Lin}, {Malacaria}, {Stewart}, \& {Wadiasingh}}]{lys+23}
{Lower}, M.~E., {Younes}, G., {Scholz}, P., {et~al.} 2023, \apj, 945, 153, \dodoi{10.3847/1538-4357/acbc7c}

\bibitem[{{Lower} {et~al.}(2024){Lower}, {Johnston}, {Lyutikov}, {Melrose}, {Shannon}, {Weltevrede}, {Caleb}, {Camilo}, {Cameron}, {Dai}, {Hobbs}, {Li}, {Rajwade}, {Reynolds}, {Sarkissian}, \& {Stappers}}]{ljl+24}
{Lower}, M.~E., {Johnston}, S., {Lyutikov}, M., {et~al.} 2024, Nature Astronomy, 8, 606, \dodoi{10.1038/s41550-024-02225-8}

\bibitem[{{Lynch} {et~al.}(2015){Lynch}, {Archibald}, {Kaspi}, \& {Scholz}}]{lak+15}
{Lynch}, R.~S., {Archibald}, R.~F., {Kaspi}, V.~M., \& {Scholz}, P. 2015, \apj, 806, 266, \dodoi{10.1088/0004-637X/806/2/266}

\bibitem[{{Maan} {et~al.}(2019){Maan}, {Joshi}, {Surnis}, {Bagchi}, \& {Manoharan}}]{mjs+19}
{Maan}, Y., {Joshi}, B.~C., {Surnis}, M.~P., {Bagchi}, M., \& {Manoharan}, P.~K. 2019, \apjl, 882, L9, \dodoi{10.3847/2041-8213/ab3a47}

\bibitem[{{Maan} \& {van Leeuwen}(2020)}]{2020ATel13560....1M}
{Maan}, Y., \& {van Leeuwen}, J. 2020, The Astronomer's Telegram, 13560, 1

\bibitem[{{Majid} {et~al.}(2020){Majid}, {Pearlman}, {Prince}, {Naudet}, \& {Bansal}}]{2020ATel13898....1M}
{Majid}, W.~A., {Pearlman}, A.~B., {Prince}, T.~A., {Naudet}, C.~J., \& {Bansal}, K. 2020, The Astronomer's Telegram, 13898, 1

\bibitem[{{Men} {et~al.}(2025){Men}, {McSweeney}, {Hurley-Walker}, {Barr}, \& {Stappers}}]{mmh+25}
{Men}, Y., {McSweeney}, S., {Hurley-Walker}, N., {Barr}, E., \& {Stappers}, B. 2025, arXiv e-prints, arXiv:2501.10528, \dodoi{10.48550/arXiv.2501.10528}

\bibitem[{{Mickaliger} {et~al.}(2018){Mickaliger}, {McEwen}, {McLaughlin}, \& {Lorimer}}]{mmm+18}
{Mickaliger}, M.~B., {McEwen}, A.~E., {McLaughlin}, M.~A., \& {Lorimer}, D.~R. 2018, \mnras, 479, 5413, \dodoi{10.1093/mnras/sty1785}

\bibitem[{{Mickaliger} {et~al.}(2012){Mickaliger}, {McLaughlin}, {Lorimer}, {Langston}, {Bilous}, {Kondratiev}, {Lyutikov}, {Ransom}, \& {Palliyaguru}}]{mml+12}
{Mickaliger}, M.~B., {McLaughlin}, M.~A., {Lorimer}, D.~R., {et~al.} 2012, \apj, 760, 64, \dodoi{10.1088/0004-637X/760/1/64}

\bibitem[{{Mitra} {et~al.}(2015){Mitra}, {Arjunwadkar}, \& {Rankin}}]{mar+15}
{Mitra}, D., {Arjunwadkar}, M., \& {Rankin}, J.~M. 2015, \apj, 806, 236, \dodoi{10.1088/0004-637X/806/2/236}

\bibitem[{{Pearlman} {et~al.}(2018){Pearlman}, {Majid}, {Prince}, {Kocz}, \& {Horiuchi}}]{pmp+18}
{Pearlman}, A.~B., {Majid}, W.~A., {Prince}, T.~A., {Kocz}, J., \& {Horiuchi}, S. 2018, \apj, 866, 160, \dodoi{10.3847/1538-4357/aade4d}

\bibitem[{{Pearlman} {et~al.}(2020){Pearlman}, {Majid}, {Prince}, {Naudet}, {Bansal}, \& {Horiuchi}}]{2020ATel13966....1P}
{Pearlman}, A.~B., {Majid}, W.~A., {Prince}, T.~A., {et~al.} 2020, The Astronomer's Telegram, 13966, 1

\bibitem[{{Pennucci} {et~al.}(2015){Pennucci}, {Possenti}, {Esposito}, {Rea}, {Haggard}, {Baganoff}, {Burgay}, {Coti Zelati}, {Israel}, \& {Minter}}]{ppe+15}
{Pennucci}, T.~T., {Possenti}, A., {Esposito}, P., {et~al.} 2015, \apj, 808, 81, \dodoi{10.1088/0004-637X/808/1/81}

\bibitem[{{Pilia} {et~al.}(2016){Pilia}, {Hessels}, {Stappers}, {Kondratiev}, {Kramer}, {van Leeuwen}, {Weltevrede}, {Lyne}, {Zagkouris}, {Hassall}, {Bilous}, {Breton}, {Falcke}, {Grie{\ss}meier}, {Keane}, {Karastergiou}, {Kuniyoshi}, {Noutsos}, {Os{\l}owski}, {Serylak}, {Sobey}, {ter Veen}, {Alexov}, {Anderson}, {Asgekar}, {Avruch}, {Bell}, {Bentum}, {Bernardi}, {B{\^\i}rzan}, {Bonafede}, {Breitling}, {Broderick}, {Br{\"u}ggen}, {Ciardi}, {Corbel}, {de Geus}, {de Jong}, {Deller}, {Duscha}, {Eisl{\"o}ffel}, {Fallows}, {Fender}, {Ferrari}, {Frieswijk}, {Garrett}, {Gunst}, {Hamaker}, {Heald}, {Horneffer}, {Jonker}, {Juette}, {Kuper}, {Maat}, {Mann}, {Markoff}, {McFadden}, {McKay-Bukowski}, {Miller-Jones}, {Nelles}, {Paas}, {Pandey-Pommier}, {Pietka}, {Pizzo}, {Polatidis}, {Reich}, {R{\"o}ttgering}, {Rowlinson}, {Schwarz}, {Smirnov}, {Steinmetz}, {Stewart}, {Swinbank}, {Tagger}, {Tang}, {Tasse}, {Thoudam}, {Toribio}, {van der Horst}, {Vermeulen}, {Vocks}, {van Weeren}, {Wijers}, {Wijnands}, {Wijnholds},
  {Wucknitz}, \& {Zarka}}]{phs+16}
{Pilia}, M., {Hessels}, J.~W.~T., {Stappers}, B.~W., {et~al.} 2016, \aap, 586, A92, \dodoi{10.1051/0004-6361/201425196}

\bibitem[{{Price}(2016)}]{2016ascl.soft03013P}
{Price}, D.~C. 2016, {PyGDSM: Python interface to Global Diffuse Sky Models}, Astrophysics Source Code Library, record ascl:1603.013

\bibitem[{{Radhakrishnan} \& {Cooke}(1969)}]{rvm_rc}
{Radhakrishnan}, V., \& {Cooke}, D.~J. 1969, \aplett, 3, 225

\bibitem[{{Rajwade} {et~al.}(2020){Rajwade}, {Stappers}, {Lyne}, {Mickaliger}, {Preston}, {Keith}, {Weltevrede}, {Kramer}, {van der Horst}, {Kouveliotou}, \& {O'Connor}}]{2020ATel13554....1R}
{Rajwade}, K., {Stappers}, B., {Lyne}, A., {et~al.} 2020, The Astronomer's Telegram, 13554, 1

\bibitem[{{Rajwade} {et~al.}(2022){Rajwade}, {Stappers}, {Lyne}, {Shaw}, {Mickaliger}, {Liu}, {Kramer}, {Desvignes}, {Karuppusamy}, {Enoto}, {G{\"u}ver}, {Hu}, \& {Surnis}}]{rsl+22}
{Rajwade}, K.~M., {Stappers}, B.~W., {Lyne}, A.~G., {et~al.} 2022, \mnras, 512, 1687, \dodoi{10.1093/mnras/stac446}

\bibitem[{{Ransom}(2001)}]{presto}
{Ransom}, S.~M. 2001, PhD thesis, Harvard University

\bibitem[{{Scholz} {et~al.}(2017){Scholz}, {Camilo}, {Sarkissian}, {Reynolds}, {Levin}, {Bailes}, {Burgay}, {Johnston}, {Kramer}, \& {Possenti}}]{scs+17}
{Scholz}, P., {Camilo}, F., {Sarkissian}, J., {et~al.} 2017, \apj, 841, 126, \dodoi{10.3847/1538-4357/aa73de}

\bibitem[{{Serylak} {et~al.}(2009){Serylak}, {Stappers}, {Weltevrede}, {Kramer}, {Jessner}, {Lyne}, {Jordan}, {Lazaridis}, \& {Zensus}}]{ssw+09}
{Serylak}, M., {Stappers}, B.~W., {Weltevrede}, P., {et~al.} 2009, \mnras, 394, 295, \dodoi{10.1111/j.1365-2966.2008.14260.x}

\bibitem[{{Stamatikos} {et~al.}(2020){Stamatikos}, {Barthelmy}, {Cummings}, {Evans}, {Krimm}, {Laha}, {Lien}, {Markwardt}, {Palmer}, {Sakamoto}, \& {Ukwatta}}]{2020GCN.27384....1S}
{Stamatikos}, M., {Barthelmy}, S.~D., {Cummings}, J.~R., {et~al.} 2020, GRB Coordinates Network, 27384, 1

\bibitem[{{Torne} {et~al.}(2015){Torne}, {Eatough}, {Karuppusamy}, {Kramer}, {Paubert}, {Klein}, {Desvignes}, {Champion}, {Wiesemeyer}, {Kramer}, {Spitler}, {Thum}, {Gusten}, {Schuster}, \& {Cognard}}]{tek+15}
{Torne}, P., {Eatough}, R.~P., {Karuppusamy}, R., {et~al.} 2015, \mnras, 451, L50, \dodoi{10.1093/mnrasl/slv063}

\bibitem[{{Torne} {et~al.}(2017){Torne}, {Desvignes}, {Eatough}, {Karuppusamy}, {Paubert}, {Kramer}, {Cognard}, {Champion}, \& {Spitler}}]{tde+17}
{Torne}, P., {Desvignes}, G., {Eatough}, R.~P., {et~al.} 2017, \mnras, 465, 242, \dodoi{10.1093/mnras/stw2757}

\bibitem[{{Torne} {et~al.}(2020){Torne}, {Liu}, {Cognard}, {Desvignes}, {Karuppusamy}, {Kramer}, {Paubert}, {Lyne}, {Rajwade}, {Stappers}, {Eatough}, {Sanchez}, {Macias-Perez}, {Ladjelate}, {Berta}, {Sanchez-Portal}, {Navarro}, {Bongiovanni}, {Kramer}, \& {Schuster}}]{2020ATel14001....1T}
{Torne}, P., {Liu}, K., {Cognard}, I., {et~al.} 2020, The Astronomer's Telegram, 14001, 1

\bibitem[{{Torne} {et~al.}(2022){Torne}, {Bell}, {Bintley}, {Desvignes}, {Berry}, {Dempsey}, {Ho}, {Parsons}, {Eatough}, {Karuppusamy}, {Kramer}, {Kramer}, {Liu}, {Paubert}, {Sanchez-Portal}, \& {Schuster}}]{tbb+22}
{Torne}, P., {Bell}, G.~S., {Bintley}, D., {et~al.} 2022, \apjl, 925, L17, \dodoi{10.3847/2041-8213/ac4caa}

\bibitem[{{van Straten} {et~al.}(2011){van Straten}, {Demorest}, {Khoo}, {Keith}, {Hotan}, \& {et al.}}]{psrchive}
{van Straten}, W., {Demorest}, P., {Khoo}, J., {et~al.} 2011, {PSRCHIVE: Development Library for the Analysis of Pulsar Astronomical Data}, Astrophysics Source Code Library, record ascl:1105.014.
\newblock \doeprint{1105.014}

\bibitem[{Virtanen {et~al.}(2020)Virtanen, Gommers, Oliphant, Haberland, Reddy, Cournapeau, Burovski, Peterson, Weckesser, Bright, {van der Walt}, Brett, Wilson, Millman, Mayorov, Nelson, Jones, Kern, Larson, Carey, Polat, Feng, Moore, {VanderPlas}, Laxalde, Perktold, Cimrman, Henriksen, Quintero, Harris, Archibald, Ribeiro, Pedregosa, {van Mulbregt}, \& {SciPy 1.0 Contributors}}]{scipy}
Virtanen, P., Gommers, R., Oliphant, T.~E., {et~al.} 2020, Nature Methods, 17, 261, \dodoi{10.1038/s41592-019-0686-2}

\bibitem[{{Wahl} {et~al.}(2022){Wahl}, {McLaughlin}, {Gentile}, {Jones}, {Spiewak}, {Arzoumanian}, {Crowter}, {Demorest}, {DeCesar}, {Dolch}, {Ellis}, {Ferdman}, {Ferrara}, {Fonseca}, {Garver-Daniels}, {Jones}, {Lam}, {Levin}, {Lewandowska}, {Lorimer}, {Lynch}, {Madison}, {Ng}, {Nice}, {Pennucci}, {Ransom}, {Ray}, {Stairs}, {Stovall}, {Swiggum}, \& {Zhu}}]{wmg+22}
{Wahl}, H.~M., {McLaughlin}, M.~A., {Gentile}, P.~A., {et~al.} 2022, \apj, 926, 168, \dodoi{10.3847/1538-4357/ac4045}

\bibitem[{{Waskom}(2021)}]{seaborn}
{Waskom}, M. 2021, The Journal of Open Source Software, 6, 3021, \dodoi{10.21105/joss.03021}

\bibitem[{{Weltevrede} {et~al.}(2006){Weltevrede}, {Edwards}, \& {Stappers}}]{wes+06}
{Weltevrede}, P., {Edwards}, R.~T., \& {Stappers}, B.~W. 2006, \aap, 445, 243, \dodoi{10.1051/0004-6361:20053088}

\bibitem[{{Yan} {et~al.}(2018){Yan}, {Wang}, {Manchester}, {Wen}, \& {Yuan}}]{ywm+18}
{Yan}, W.~M., {Wang}, N., {Manchester}, R.~N., {Wen}, Z.~G., \& {Yuan}, J.~P. 2018, \mnras, 476, 3677, \dodoi{10.1093/mnras/sty470}

\bibitem[{{Yao} {et~al.}(2017){Yao}, {Manchester}, \& {Wang}}]{ymw17}
{Yao}, J.~M., {Manchester}, R.~N., \& {Wang}, N. 2017, \apj, 835, 29, \dodoi{10.3847/1538-4357/835/1/29}

\bibitem[{{Zheng} {et~al.}(2017){Zheng}, {Tegmark}, {Dillon}, {Kim}, {Liu}, {Neben}, {Jonas}, {Reich}, \& {Reich}}]{ztd+17}
{Zheng}, H., {Tegmark}, M., {Dillon}, J.~S., {et~al.} 2017, \mnras, 464, 3486, \dodoi{10.1093/mnras/stw2525}

\end{thebibliography}
